\documentclass[preprint,prd,nofootinbib,tightenlines,amsmath]{revtex4}
\usepackage{enumerate}
\usepackage{multirow}

\usepackage{epsfig}
\usepackage{graphics,color}      
\usepackage{verbatim}
\usepackage{amsmath}   
\catcode`\@=11
\def\sla@#1#2#3#4#5{{%
 \setbox\z@\hbox{$\m@th#4#5$}%
 \setbox\tw@\hbox{$\m@th#4#1$}%
 \dimen4\wd\ifdim\wd\z@<\wd\tw@\tw@\else\z@\fi
 \dimen@\ht\tw@
 \advance\dimen@-\dp\tw@ \advance\dimen@-\ht\z@
 \advance\dimen@\dp\z@
 \divide\dimen@\tw@ \advance\dimen@-#3\ht\tw@
 \advance\dimen@-#3\dp\tw@ \dimen@ii#2\wd\z@
 \raise-\dimen@\hbox to\dimen4{%
 \hss\kern\dimen@ii\box\tw@\kern-\dimen@ii\hss}%
 \llap{\hbox to\dimen4{\hss\box\z@\hss}}}}

\def\slashed#1{%
 \expandafter\ifx\csname sla@\string#1\endcsname{\rm ~Re}lax
{\mathpalette{\sla@/00}{#1}}
\fi}
\def\declareslashed#1#2#3#4#5{%
 \expandafter\def\csname sla@\string#5\endcsname{%
#1{\mathpalette{\sla@{#2}{#3}{#4}}{#5}}}}
 \catcode`\@=12
\declareslashed{}{/}{.08}{0}{D}
 \declareslashed{}{/}{.1}{0}{A}
 \declareslashed{}{/}{0}{-.05}{k}
 \declareslashed{}{/}{.1}{0}{\partial}
 \declareslashed{}{\not}{-.6}{0}{f}

\def\lsim{\mathrel {\vcenter {\baselineskip 0pt \kern 0pt
    \hbox{$<$} \kern 0pt \hbox{$\sim$} }}}
\def\gsim{\mathrel {\vcenter {\baselineskip 0pt \kern 0pt
    \hbox{$>$} \kern 0pt \hbox{$\sim$} }}}

\begin{document}

\baselineskip=15pt
\preprint{}

\title{CP violating anomalous couplings in $W$ jet production at the LHC}

\author{Hai Tao Li$^{1,2}$\footnote{Electronic address: haitao.li@monash.edu},  German Valencia$^{2}$\footnote{Electronic address: german.valencia@monash.edu }}

\affiliation{$^1$ ARC Centre of Excellence for Particle Physics at the Terascale}

\affiliation{$^2$ School of Physics and Astronomy, Monash University, VIC-3800 Australia}

\date{\today}

\vskip 1cm
\begin{abstract}

T-odd correlations in $Wj$ production at the LHC have been studied recently as a way to measure a phase produced by QCD at NLO by Frederix {\it et. al.} \cite{Frederix:2014cba}. That study found that the induced asymmetry could be observed with 20 fb$^{-1}$ of 8 TeV data. These T-odd asymmetries can also be induced by CP violating new physics interfering with the SM at LO. In this paper we study this possibility using effective Lagrangians to describe the new physics. We find that the leading contribution arises at dimension eight, and that it necessarily introduces flavor changing neutral currents as well. We discuss the constraints that can be placed on the flavor structure of the new physics operator from studies of FCNC in kaon and B meson decays and then compare the T-odd correlations in $Wj$ induced by CP violating new physics to those induced by QCD at NLO. We quantify the level at which these couplings can be probed at the LHC, and find that they will not affect a measurement of the NLO QCD phases.

\end{abstract}

\pacs{PACS numbers: }

\maketitle
    

\newpage

\section{Introduction}

The Large Hadron Collider (LHC) is well on its way to testing the standard model (SM) in detail. As part of this program, the couplings between known particles are parametrized in a general form to quantify possible deviations from their standard model (SM) values, aka the study of `anomalous couplings'. One example is the $WW\gamma$ triple gauge boson coupling which can be tested by studying $W\gamma$ production. In this case the $W\gamma$ rate measured by ATLAS and CMS is compared to the most accurate prediction existing for the SM (at NLO) and any discrepancy is interpreted as a quantitative limit on the anomalous couplings that may affect this vertex. In addition to the total rate, the $W\gamma$ process permits the measurement of T-odd correlations. These are special kinematic correlations that can only be induced by CP violation or by unitarity phases that can occur at NLO and can be significant only in the case of QCD.

A very similar situation occurs with the process $Wj$, where the LHC has already published results for 4.6~fb$^{-1}$ at 7 TeV \cite{Aad:2014qxa}, for 19.6~fb$^{-1}$ at 8 TeV \cite{Khachatryan:2016fue} and for 2.2~fb$^{-1}$ at 13 TeV \cite{Sirunyan:2017wgx} that include measurements of both cross-section and angular correlations between the jets and the lepton from $W$ decay. Beyond the standard model (BSM) it is possible to modify this process and, in particular, to introduce CP violating contributions that would produce T-odd correlations involving the beam, the jet and the lepton momenta. 

This possibility was studied in the context of the Tevatron in Ref.~\cite{Dawson:1995wg,Hagiwara:2006qe} where it was noted that the $p\bar{p}$ initial state would allow one to isolate CP odd effects by comparing the $W^+j$ and $W^-j$ processes. This is no longer possible at LHC where the initial $pp$ state results in a larger cross-section for $W^+j$. In this situation the T-odd correlations can also be produced by discontinuities that occur at loop order. This effect is interesting in its own right and has been discussed recently  \cite{Frederix:2014cba}, concluding that 20 fb$^{-1}$ of 8 TeV LHC data would suffice to measure it. In this paper we will compare the T-odd correlations that can be induced by CP violating new physics to the aforementioned QCD effect.

\begin{figure}[thb]
\includegraphics[width=.65\textwidth]{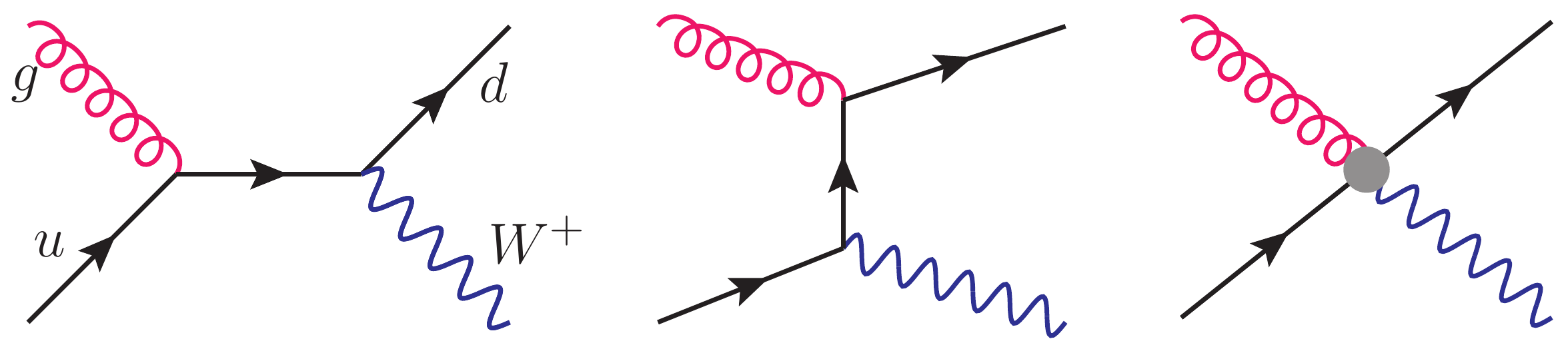}
\caption{Feynman diagrams  contributing to $Wj$ production. The two on the left are the SM diagrams whose vertices can be modified by the new physics. On the right a seagull diagram possibly induced by higher dimension operators.
\label{feyn-dia}}
\end{figure}

At leading order, the process $pp\to W^+ j$ proceeds via the two diagrams on the left of Figure~\ref{feyn-dia}. To include BSM effects in a model independent way it is convenient to use an effective Lagrangian which respects the symmetries of the SM, and which encodes the leading effects of the new physics in operators of dimension six or higher. New CP violation occurs as phases in the couplings appearing in these operators. Generally speaking, these operators can modify the $qqG$ and $UDW$ couplings appearing in Figure~\ref{feyn-dia} introducing a CP phase. However, this is not enough to generate a T-odd correlation in the process $pp\to W^+j$. Typically this requires the existence of a different amplitude such as the `seagull', shown also in Figure~\ref{feyn-dia} which can also be produced  by  the higher dimension operators \cite{Dawson:1995wg}.

\section{Effective Lagrangian for Physics BSM}

\subsection{Operators at dimension six}

A list of operators of dimension six that are consistent with all the symmetries of the SM exists in the literature \cite{Buchmuller:1985jz,Grzadkowski:2010es}. Amongst this list there are several   operators that modify the charged current interaction and that could induce CP violation. In principle CP violation can be encoded in a complex coupling for any of the operators in the list. However, our study requires more than that. The CP-odd observables will consist of (naive)~T-odd triple products that, when due to CP violation, arise through interference with the SM tree-level amplitude. Since $Wj$ production at LHC involves mainly light quarks we need a CP violating operator that interferes with the SM without helicity suppression. This last requirement eliminates all operators with scalar or right-handed quark bi-linears. 

In the notation of \cite{Buchmuller:1985jz} we are left with the following possibilities:  ${\cal O}^1_{\Phi Q}$, ${\cal O}^3_{\Phi Q}$ and  ${\cal O}_{qW}$ which modify the left-handed charged current, and in addition, with ${\cal O}_{qG}$ which introduces a $udWG$ vertex that could also produce the desired effect. These operators are (in order)
\begin{eqnarray}
{\cal L} &\supset & \frac{i a_{ij}}{\Lambda^2}  \underbrace{  (\Phi^\dagger D_\mu \Phi) (\bar{q}_i\gamma^\mu q_j)}_{{\cal O}_{\Phi q}^1} + \frac{i b_{ij}}{\Lambda^2} \underbrace{(\Phi^\dagger D_\mu \tau^I\Phi) (\bar{q}_i\gamma^\mu \tau^Iq_j)}_{{\cal O}_{\Phi q}^3}  \nonumber \\
&+& \frac{i c_{ij}}{\Lambda^2} \underbrace{\bar{q}_i\tau^I\gamma_\mu D_\nu q_j W^{I\mu\nu}}_{{\cal O}_{qW}} +
\frac{i d_{ij}}{\Lambda^2}\underbrace{\bar{q}_i T^A\gamma_\mu D_\nu q_j G^{A\mu\nu}}_{{\cal O}_{qG}} + h.c. ,
\label{operators}
\end{eqnarray}
where we have explicitly written generation indices $i,j$; $q$ stands for a left-handed quark doublet; and $\Phi$ is the SM scalar doublet. In order for any phase in the $a_{ij}$, $b_{ij}$, $c_{ij}$ or $d_{ij}$ couplings to generate a CP-odd interference, the diagrams of Figure~\ref{feyn-dia} cannot have the same phase. It is evident that the first two operators will not produce a seagull diagram, and that they will produce the same phase in the other two diagrams. This implies they cannot introduce CP-odd asymmetries into the $pp\to Wj$ process. The operator ${\cal O}_{qG}$ produces amplitudes that vanish for on-shell gluons and hence does not contribute to the process of interest at leading order. 

We are left with ${\cal O}_{qW}$ which turns out not to contribute either, as we have verified with an explicit calculation, although in this case the reason is not obvious. The explicit calculation shows that all the diagrams arising from SM + ${\cal O}_{qW}$ have the same phase so no CP violating interference is possible. One way to understand this is to look at the operator list in \cite{Grzadkowski:2010es} where this operator does not occur. Use of the equations of motion and integration by parts replace this operator with operators of the form ${\cal O}^1_{\phi q}$, 
${\cal O}^3_{\phi q}$, $ \bar{q}_i \sigma_{\mu\nu}\tau^I d_j W^{I\mu\nu} $ and $\bar{q}\gamma_\mu q\bar\ell \gamma^\mu\ell$ where $\ell$ is the left handed lepton doublet. As already mentioned, the first two do not introduce CP violation in $Wj$ production; and the second one has an interference with the SM that vanishes for massless quarks. The operator ${\cal O}_{qW}$ is thus equivalent to the four fermion form  $\bar{q}\gamma_\mu q\bar\ell \gamma^\mu\ell$ for our purposes. It is now obvious that this has the exact same structure as the SM amplitude and any phase in its coupling will disappear from the observables.

We are thus led  to the same conclusion as Ref.~\cite{Dawson:1995wg}: the NP operator must give rise to a seagull diagram with a non-zero phase different from that of the other two diagrams  while maintaining the left-handed nature of the SM amplitude, and this is only possible through vertices that depend on the fermion momentum. These are first generated at dimension eight.

\subsection{Operators at dimension eight}

To construct the relevant dimension eight operators we consider the different ingredients that are needed while taking into account  our findings from the previous section:
\begin{itemize}
\item The quark fields must enter as a left-handed bilinear so that the interference with the SM amplitude is not suppressed by small quark masses. 
\item The operator must contain quark momenta so that the phases of the different diagrams can be different. These first two restrictions leave us to consider only bilinears with the forms
\begin{eqnarray}
D_\alpha \bar{q}_i \gamma_\mu D_\beta q_j,\ D_\alpha \bar{q}_i \gamma_\mu \tau^I D_\beta q_j
\end{eqnarray}
\item At leading order in QCD, there will only be one jet so the parton content of the operator needs $qqg$. The field strength tensor $G_{\mu\nu}$ does not contribute to an amplitude with an on-shell gluon, so the gluon must come from one of the covariant derivatives acting on the quark fields.
\item The SM amplitude produces the lepton pair from an on-shell $W$ and therefore its interference with a new physics operator will suffer a very large kinematical suppression unless the NP operator results in $m_{\ell \nu} \sim M_W$. This rules out lepton bilinears from appearing explicitly in the operator. In other words, the operator should produce $q\bar{q}\to Wg$ and not just $q\bar{q}\to \ell\nu g$.
\item This requirement of an explicit $W$ field can be satisfied with an elementary Higgs boson, with two possibilities
\begin{eqnarray}
\Phi^\dagger {D}_\mu \tau^I\Phi, \ (D^\mu W_{\nu\mu})^I
\end{eqnarray}
which are equivalent through use of the equations of motion. 
\end{itemize}

From this discussion, and after symmetrizing the scalar bilinear for convenience, it should be clear that the effective operator required can only have the form
\begin{eqnarray}
{\cal O}_8&\sim & i\frac{g_{ij}}{\Lambda^4} D_\alpha \bar q_i \gamma_\mu \tau^I D^\alpha q_j \ (\Phi^\dagger \overleftrightarrow{D}_\mu \tau^I\Phi) +h.c.
\label{dim8un}
\end{eqnarray}
One might think that a rearrangement of Lorentz indices leads to a second form, but it is easy to see that this second form produces amplitudes proportional to the light quark masses by using the equation of motion.

Eq.~\ref{dim8un} still contains multipole operators corresponding to the different flavor structures. In order to violate CP,  the couplings $g_{ij}$ must have a non-zero phase. Explicitly adding the hermitian conjugate of the operator we  have
\begin{eqnarray}
{\cal L}_8 
&=&  \frac{ \left(g_{ij}+g^*_{ji}\right) }{\Lambda^4} i D_\alpha \bar q_i \gamma_\mu \tau^I D^\alpha q_j \ (\Phi^\dagger \overleftrightarrow{D}_\mu \tau^I\Phi).
\label{opweuse}
\end{eqnarray}
This last result shows that it is not possible to have CP violation with $i=j$, and therefore, that CP violation in these operators is necessarily accompanied by FCNC. This leaves operators with  the 1-2, 1-3 or 2-3 family structure. In our numerical study we will confirm what is evident: the 1-2 operator is the dominant one even though it has the strictest existing low energy constraints. 

We could also write similar operators for an effective theory appropriate for a composite Higgs \cite{Alonso:2012px,Buchalla:2013rka,Gavela:2014vra} replacing the factor $(\Phi^\dagger D_\mu \tau^I\Phi) $ with a factor $\langle L_\mu U \tau^I U^\dagger\rangle$ in the notation of Ref.~\cite{Buchalla:2013rka}. In this study, we will proceed with the elementary Higgs effective theory.
\subsection{FCNC}

\begin{figure}[thb]
\includegraphics[width=.65\textwidth]{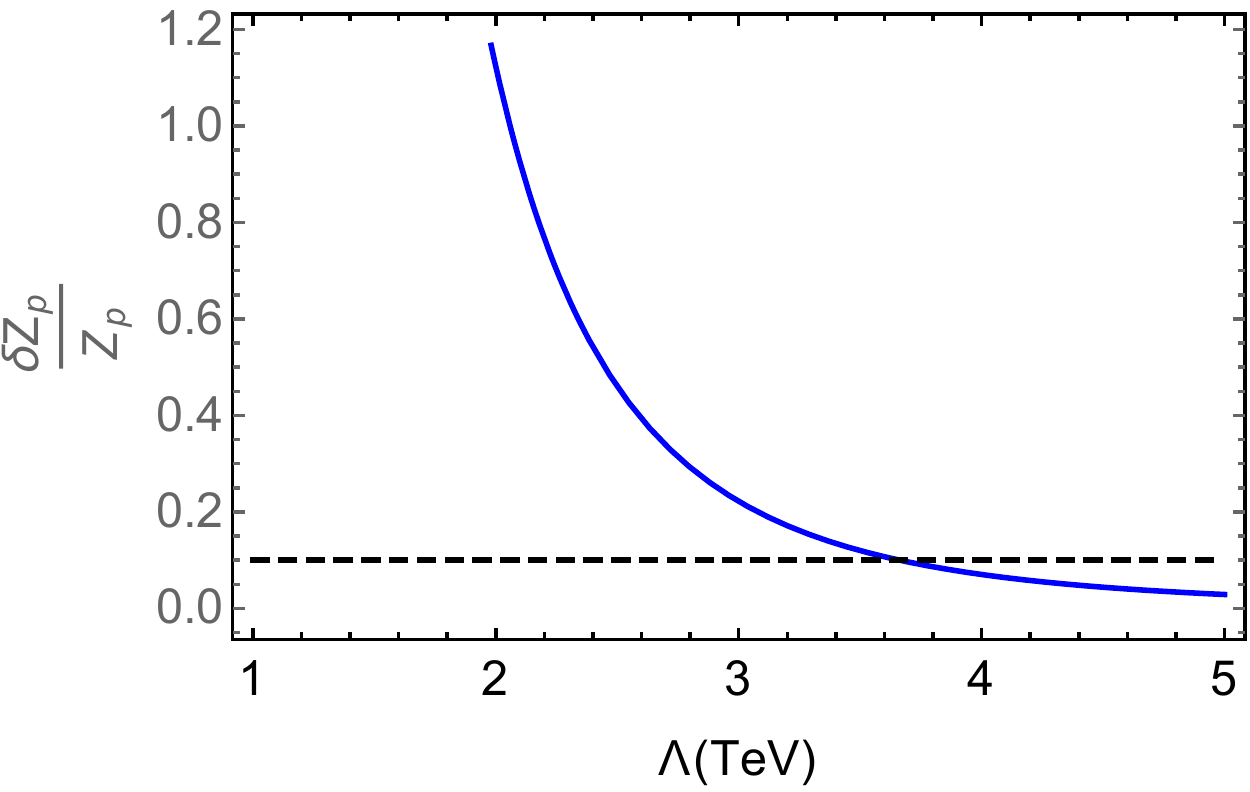}
\caption{Correction to the SM $Z$-penguin amplitude, $Z_p$, as a function of $\Lambda$ for $g_{12}=1, g_{21}=0$.
\label{pen-rat}}
\end{figure}

Since the CP violating component of the operator of Eq.~(\ref{opweuse}) only occurs in conjunction with FCNC, we estimate the corresponding existing constraints from low energy observables. For this purpose we choose the operator involving the first and second generations, $g_{1,2}\neq 0$.  The leading order FCNC vertices consist of $u\leftrightarrow c$ and  $s\leftrightarrow d$ couplings to $Z$:
\begin{eqnarray}
\bar{d}sZ:&& i\frac{g v^2}{8 c_W}\frac{p_d\cdot p_s}{\Lambda^4} \left( (g_{12}+g^*_{21}) V_{cs}V^*_{ud}+(g_{21}+g^*_{12}) V_{us}V^*_{cd} \right)\ \bar{d} \gamma_\mu P_L \ s \nonumber \\
\bar{u}cZ:&& -i\frac{gv^2}{8 c_W}\frac{p_c\cdot p_u }{\Lambda^4}\left(g_{12}+g^*_{21}\right)\ \bar{u} \gamma_\mu P_L \ c
\label{gpens}
\end{eqnarray}
and their hermitian conjugates. In Eq.~(\ref{gpens}) $s_W,c_W$ stand for the sine and cosine of the Weinberg angle respectively, and $v=246$~GeV for the usual Higgs vacuum expectation value. The most stringent constraints on these vertices arise from kaon physics and we can obtain a simple estimate by comparing the first one to the usual $Z$-penguin diagram that mediates FCNC in that case. 
In the notation of \cite{Buras:1992uf}, the one-loop SM $Z$-penguin is given by
\begin{eqnarray}
\bar{s}dZ:&&V_{td}V_{ts}^* \frac{G_F}{\sqrt{2}}\frac{e}{\pi^2}M_Z^2\frac{c_W}{s_W}C(x_t)\ \bar{s} \gamma_\mu P_L \ d
\label{bupen}
\end{eqnarray}
where  $C(x)$ is an Inami-Lim loop function \cite{Inami:1980fz} which is approximately equal to  $ 0.85$  for $x_t=m_t^2/m_W^2$~. In Figure~\ref{pen-rat} we plot the correction to the SM $Z$-penguin amplitude, Eq.~(\ref{gpens}), as a fraction of the  SM $Z$-penguin amplitude, Eq.~(\ref{bupen}), as a function of the new physics scale $\Lambda$ for $g_{12}=1, g_{21}=0$. We see that a scale as low as 3.5~TeV would only modify the SM amplitude by 10\%. The only kaon process without long distance uncertainties that has been measured so far is $K^+\to \pi^+ \nu \bar\nu$, and the BNL-787 (and 949) results \cite{expres} still have enough uncertainty to allow a NP contribution as large as the SM one \cite{Olive:2016xmw}. Conservatively we conclude from Figure~\ref{pen-rat} that the current bound from kaon decay is a relatively weak $\Lambda \gsim 2$~TeV.

Had we chosen instead operators involving the second and third generations,  $g_{2,3}\neq 0$,  Eq.~(\ref{opweuse}) would then generate the penguin
\begin{eqnarray}
\bar{b}sZ:&& i \frac{g v^2}{8 c_W}\frac{p_b\cdot p_s }{\Lambda^4}\left((g^*_{23}+g_{32})V_{cs}V^*_{tb}+(g_{23}+g^*_{32})V_{ts}V^*_{cb}\right)\ \bar{b} \gamma_\mu P_L \ s 
\label{gpenb}
\end{eqnarray}
which contributes to rare B decay. To quantify the existing bounds in this case, we note that Eq.~(\ref{gpenb}) 
modifies two terms in the effective weak Hamiltonian relevant for $b$ decay
\begin{eqnarray}
{\cal H}&\supset -\frac{4G_F}{\sqrt{2}}V_{tb}V^*_{ts}\frac{e^2}{16\pi^2}\left( C_9 \bar{s} \gamma_\mu P_L \ b \bar\ell\gamma^\mu\ell +C_{10} \bar{s} \gamma_\mu P_L \ b \bar\ell\gamma^\mu\gamma_5\ell \right).
\end{eqnarray}
with respective contributions to the Wilson coefficients $C_{9,10}$ given by,
\begin{eqnarray}
C_{9}^{NP} \sim (g^*_{23}+g_{32}) \left(\frac{460~{\rm GeV}}{\Lambda}\right)^4,&& C_{10}^{NP} \sim (g^*_{23}+g_{32})\left(\frac{880~{\rm GeV}}{\Lambda}\right)^4.
\end{eqnarray}
Since the current results in B-physics actually prefer these deviations from the SM to be of order one \cite{Descotes-Genon:2015uva}, the constraints in this case are much weaker. At the same time, these operators contribute less to LHC observables in $pp\to Wj$ because they would be initiated from a charm-quark rather than light quarks. Similarly the 1-3 operator has an interference with the SM that is suppressed by $V_{ub}$.

Eq.~(\ref{opweuse}) is such that it generates CP violating neutral current vertices $d_i\bar{d}_j ZG$ or $u_i\bar{u}_j ZG$ only for $i\neq j$. These new vertices, therefore, cannot interfere with SM $Zj$ production and cannot generate CP-odd observables in $pp\to Zj$.

Of course, it is important to study different observables produced by the same type of new physics because in the most general case they give complementary information. From this perspective an LHC study that can constrain the operator of Eq.~(\ref{opweuse}) is interesting regardless of any low energy constraints from FCNC.

\section{T-odd correlations in $pp \to Wj$}

There are several T-odd correlations that can appear in $pp \to Wj$ and can be used to look for CP violation (or for strong phases beyond LO). In terms of observable momenta the basic ones are two which coincide at LO: 
\begin{align}
{\cal O}_{W}     &= \frac{2  \left( {\vec p}_j \times {\vec p}_{\rm beam} \right) \cdot  {\vec p}_\ell}{
|\left( {\vec p}_j \times {\vec p}_{\rm beam} \right)| m_W } ~,
  \nonumber \\  
{\cal O}_{T} &=  \frac{2  \left(  {\vec p}_{\rm beam} \times \vec{q}_{\perp} \right) \cdot  {\vec p}_\ell}{|\left( {\vec p}_{\rm beam} \times \vec{q}_{\perp} \right)| m_W } .
\label{eq:owot}
\end{align} 
In these expressions $\vec{q}_\perp$ is the reconstructed $W$ transverse momentum and we choose ${\vec p}_{\rm beam}$  along the $z$ axis. The observable  ${\cal O}_{T}$ is the same one proposed in Ref.~\cite{Frederix:2014cba} to measure the QCD phase and dubbed $x_\perp$ in that reference. Since the LHC is a proton-proton collider, the symmetry of the initial state would cause  ${\cal O}_{W} $ and ${\cal O}_T$ to vanish due to the ambiguity in the direction of $\vec{p}_{\rm beam}$. To circumvent this issue, Ref.~\cite{Frederix:2014cba} introduced a cut in $\Delta \eta \equiv \eta_\ell -\eta_j$ which allows an unambiguous definition of the $z$-axis.

Another possibility, which we use here, is to make these correlations even under the interchange of the two proton (beam) momenta. For this purpose we explicitly introduce an additional factor as was done for the study of CP violation in $W\gamma$ production \cite{Dawson:2013owa}. The  T-odd correlations with the extra factor are given by
\begin{align}
{\cal O}_j &=   \frac{ {\vec p}_j \cdot {\vec p}_{\rm beam}}{|{\vec p}_j \cdot {\vec p}_{\rm beam} |} {\cal O}_{W} ~, \qquad
{\cal O}_\ell =  \frac{{\vec p}_\ell \cdot {\vec p}_{\rm beam}} {|{\vec p}_\ell \cdot {\vec p}_{\rm beam} |} {\cal O}_{W} ~,
  \nonumber \\ 
{\cal O}^{T}_j &=   \frac{ {\vec p}_j \cdot {\vec p}_{\rm beam}}{|{\vec p}_j \cdot {\vec p}_{\rm beam} |} {\cal O}_{T} ~,\qquad
{\cal O}^{T}_\ell =  \frac{{\vec p}_\ell \cdot {\vec p}_{\rm beam}} {|{\vec p}_\ell \cdot {\vec p}_{\rm beam} | } {\cal O}_{T}   \, .
\label{eq:tripprods}
\end{align}
Each of these operators in Eq.~(\ref{eq:tripprods}) can then be used to construct integrated (counting) asymmetries defined as
\begin{align}
\Delta \sigma  &\equiv  \sigma\left({\cal O}^{(T)}_{j,\ell} >0\right)-\sigma\left({\cal O}^{(T)}_{j,\ell} <0 \right)~,
\nonumber \\ 
{\cal A}^{(T)}_{j,\ell} &\equiv \frac{\sigma\left({\cal O}^{(T)}_{j,\ell} >0\right)-\sigma\left({\cal O}^{(T)}_{j,\ell} <0 \right)}{\sigma\left({\cal O}^{(T)}_{j,\ell} >0\right)+\sigma\left({\cal O}^{(T)}_{j,\ell} <0\right)}~.
\label{eq:astriprods}
\end{align}
${\cal A}^{(T)}_{j,\ell}$ is the same as $\Delta \sigma$ but normalized to the cross-section. 
To illustrate our results numerically, we choose  $g_{1,2}\neq 0$ as the only non-vanishing coupling in Eq.~(\ref{opweuse}).  In this case the diagrams of Figure~\ref{feyn-dia} have $U=u,c$ and $D=d,s$ as we neglect the CKM mixing between first and third generations.

We mentioned before that at the LHC it is not realistic to compare the two processes $W^\pm j$ in order to differentiate CP violation from NLO QCD phases through T-odd asymmetries. The new physics operators could provide a different handle to isolate CP violation if they also induced T-odd correlations in the process $pp \to Zj$. In this case,  at least in principle, one can construct a truly CP odd correlation such as
\begin{eqnarray}
{\cal O}_Z =  {\vec p}_{\rm beam} . \left({\vec p}_{\ell ^+}  \times {\vec p}_{\ell^-}\right)  {\vec p}_{\rm beam} . \left({\vec p}_{\ell^+}  - {\vec p}_{\ell^-}\right). 
\label{eq:trulycp}
\end{eqnarray}
In order to induce this correlation at leading order, however, one needs interference between the tree-level SM and the new physics amplitude and this does not happen for Eq.~(\ref{opweuse}) because its neutral currents necessarily violate flavor conservation. 

\section{Numerical results}

In this section we measure the T-odd correlations  in Monte-Carlo generated samples simulating the LHC at 13~TeV. We do this for two cases: first we calculate the distributions for the SM at NLO which were studied in Ref.~\cite{Frederix:2014cba}; then we consider the tree-level SM with additional NP represented by the dimension eight operator of Eq.~(\ref{opweuse}). This will allow us to compare asymmetries induced by CP violation to those that arise from the absorptive part of QCD loops.

The SM NLO calculations are fully implemented in MG5\_aMC@NLO \cite{Alwall:2014hca} so computing this part is straightforward. The only technical difficulty here is that very large samples are required in order to extract the signal. In order to estimate the LHC sensitivity to the NP encoded in the phase and coupling $g_{12}$,  we implement the operator  Eq.~(\ref{opweuse}) in FEYNRULES \cite{Christensen:2008py,Degrande:2011ua} to generate a Universal Feynrules Output (UFO) file, and then feed this UFO file into MG5\_aMC@NLO \cite{Alwall:2014hca}. Since the T-odd correlations in the case of CP violation arise only from  the interference between the LO-SM and the NP we isolate events generated from this interference with appropriate MG5 syntax.\footnote{ The interference terms can be selected out by ``generate p p $>$ e$+$ ve j NP$\land$ 2$==$1" where NP is the order of the NP coupling. }

For both event samples,  NLO-SM and LO-SM+NP events, we use the CT14nlo pdf~\cite{Dulat:2015mca},  and apply the following selection cuts
\begin{align}
    p_{T}({\mathrm{jet}})>30~\mathrm{GeV}~, \qquad   \left|\eta({\mathrm{jet}})\right|&<4.4~,\qquad   p_{T}(\mathrm{lepton})>25~\mathrm{GeV}~, \nonumber \\
    \left|\eta({\mathrm{lepton}})\right|<2.4~, \qquad \slash{}\!\!\!\!E_{T}>25~\mathrm{GeV}~, & \qquad M_T>60~\mathrm{GeV}~, \qquad  p_{T}(W^+)>30~\mathrm{GeV}  
    \label{eq:cuts}
\end{align}
with the transverse mass defined as $M_T=\sqrt{2\left( p_{T}^\ell \slash{}\!\!\! p_{T}- \vec{p}_{T}^{~\ell}\cdot \slash{}\!\!\! \vec{p}_{T}   \right) }$~. 

\begin{figure}[thb]
\includegraphics[width=.65\textwidth]{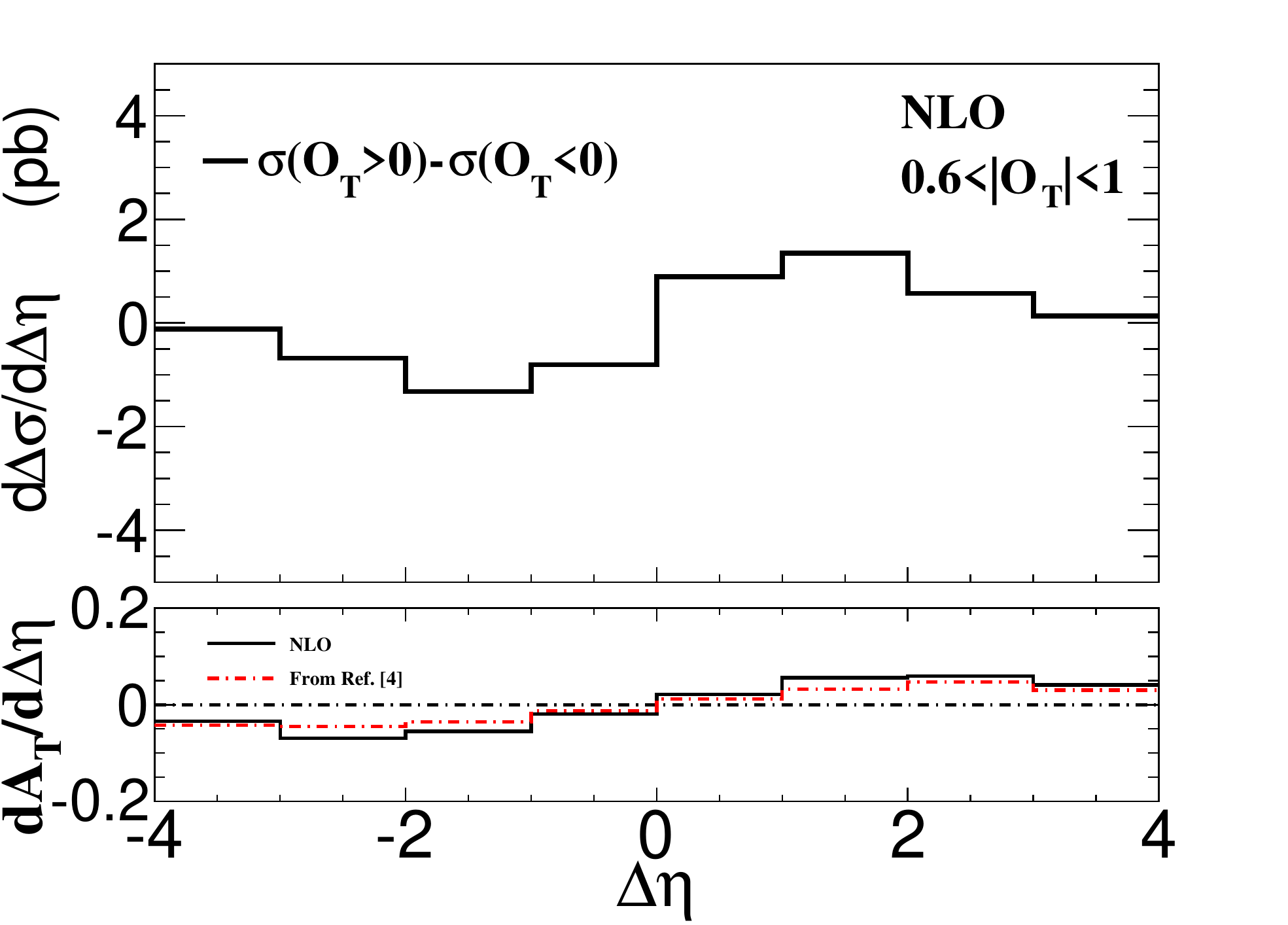}
\caption{The SM-NLO ${\cal O}_T$ distributions with the cuts in Eq.~(\ref{eq:cuts}) and $\Delta \eta > 1$ for 8~TeV compared with the results of Ref.~\cite{Frederix:2014cba}~.
\label{fig:NLO_comp}}
\end{figure}

To validate our study we first repeat the NLO-SM calculation at 8 TeV and compare our results to those in \cite{Frederix:2014cba}. We show this comparison in Figure~\ref{fig:NLO_comp}. Our parton level results are slightly larger than the results of that reference, as expected, because the latter include parton showering and detector simulation. 

\begin{figure}[thb]
\includegraphics[width=.45\textwidth]{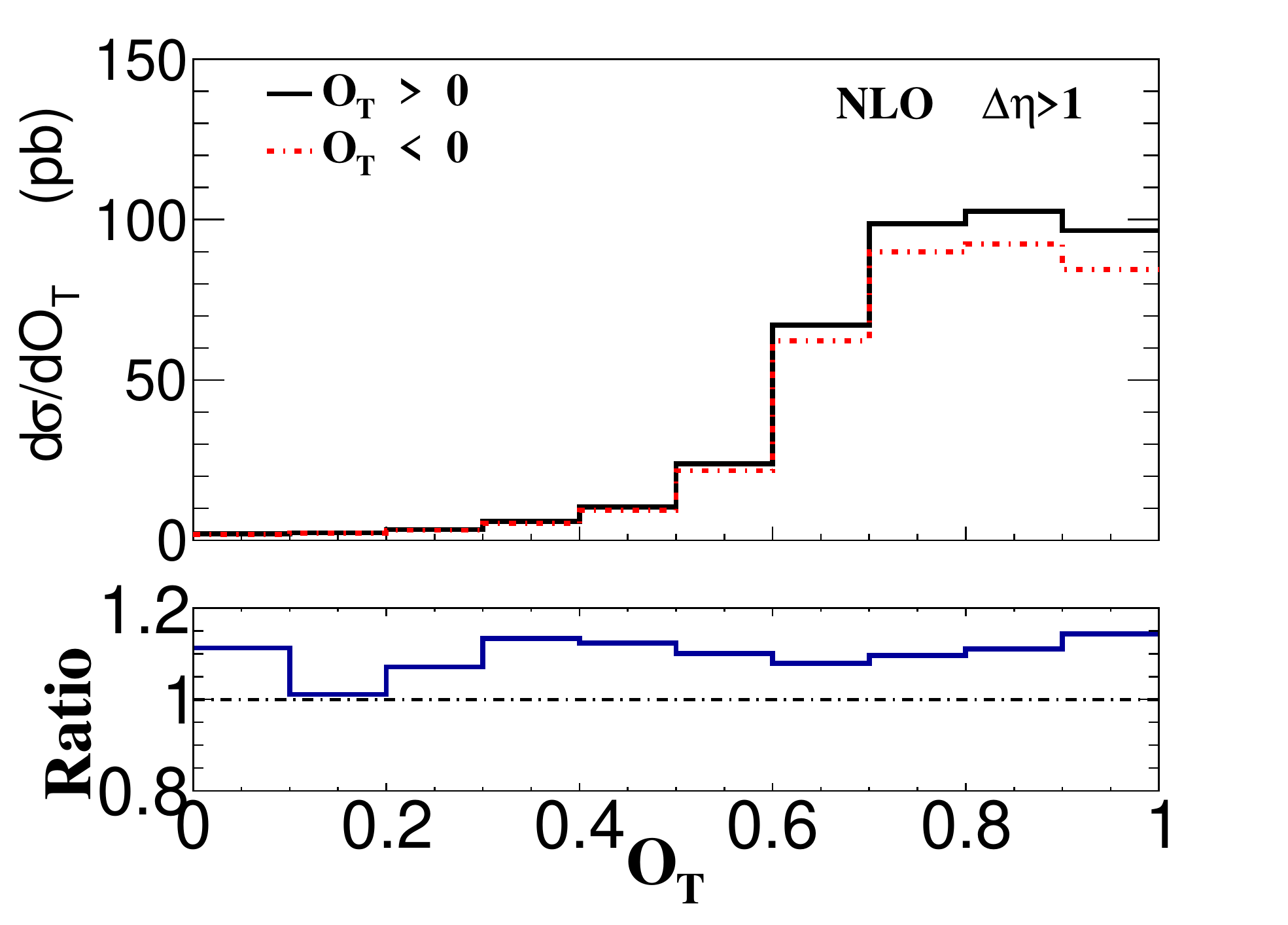}
\includegraphics[width=.45\textwidth]{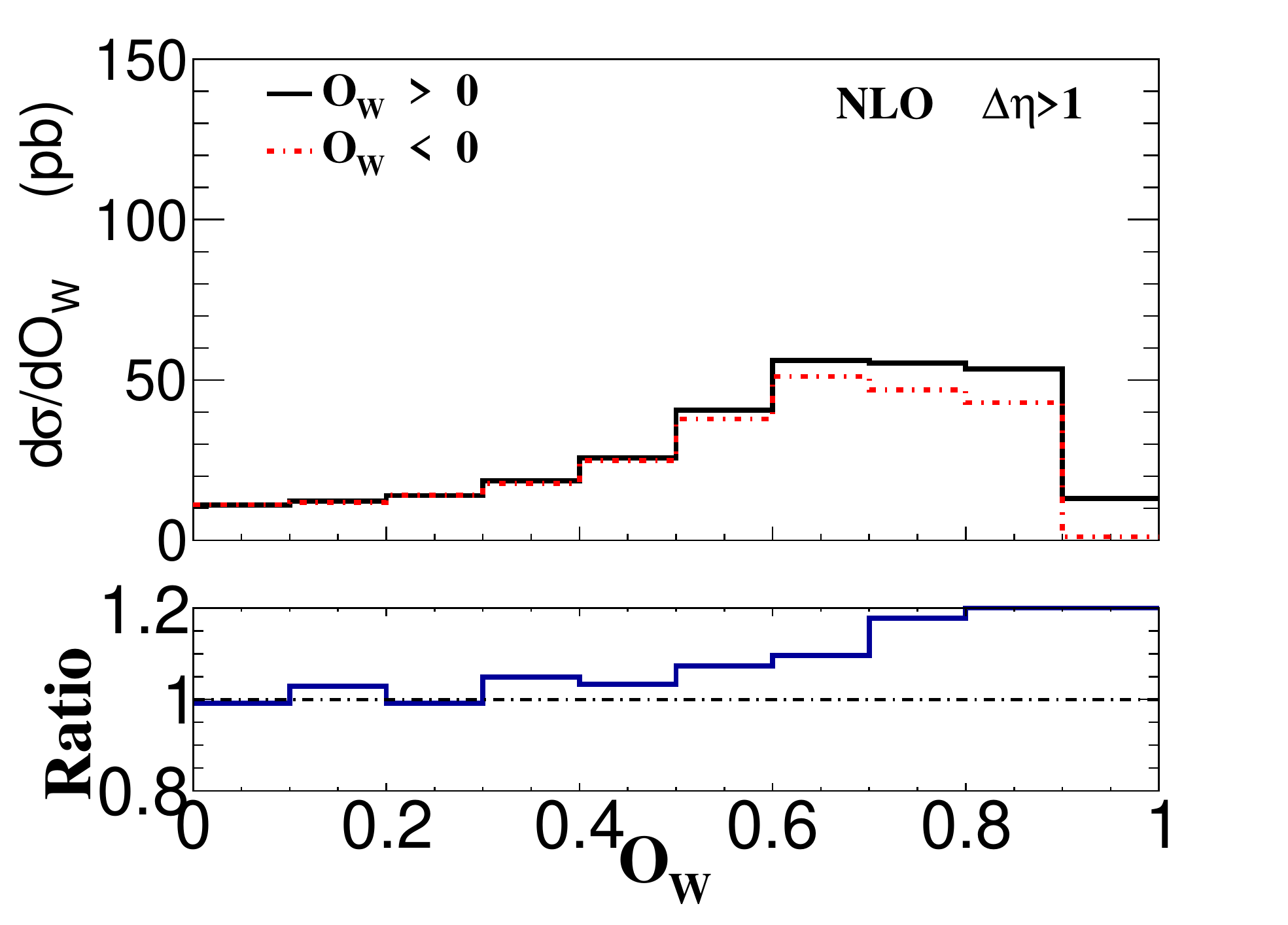}
\caption{The SM-NLO ${\cal O}_T$ and ${\cal O}_W$ distributions with the cuts in Eq.~(\ref{eq:cuts}) and $\Delta \eta > 1$ at 13~TeV.
\label{fig:NLO_otow}}
\end{figure}

We next simulate events at 13~TeV and begin by considering the correlations $\mathcal{O}_{T}$ and $\mathcal{O}_{W}$ in Eq.~(\ref{eq:owot}) which are odd functions of $\Delta \eta $~\cite{Frederix:2014cba} and which vanish when integrated over the entire phase space. As previously mentioned, the vanishing of the asymmetries over the full phase space is due to the ambiguity in defining $\vec{p}_{\rm beam}$, and this is addressed by  imposing the cut $\Delta \eta >1 $. At LO, we have that ${\cal O}_{T}={\cal O}_{W}$ because there is only one jet in the final state and $\vec{q}_\perp =- \vec{p}_{j,\perp}$, but they can be very different at higher order.  In particular, at NLO the real corrections can produce two jets in the final state. In this kinematic configuration $\mathcal{O}_{W}$ can be larger than one, and in the region where $\mathcal{O}_{W}>1$ there is no asymmetry. For this reason, imposing the cut $|\mathcal{O}_{W}|<1$  reduces the cross section but increases  the asymmetry ratio  for  $\mathcal{O}_{W}$, but not for $\mathcal{O}_{T}$, as shown in Fig.~\ref{fig:NLO_otow}. 

\begin{figure}[thb]
\includegraphics[width=.45\textwidth]{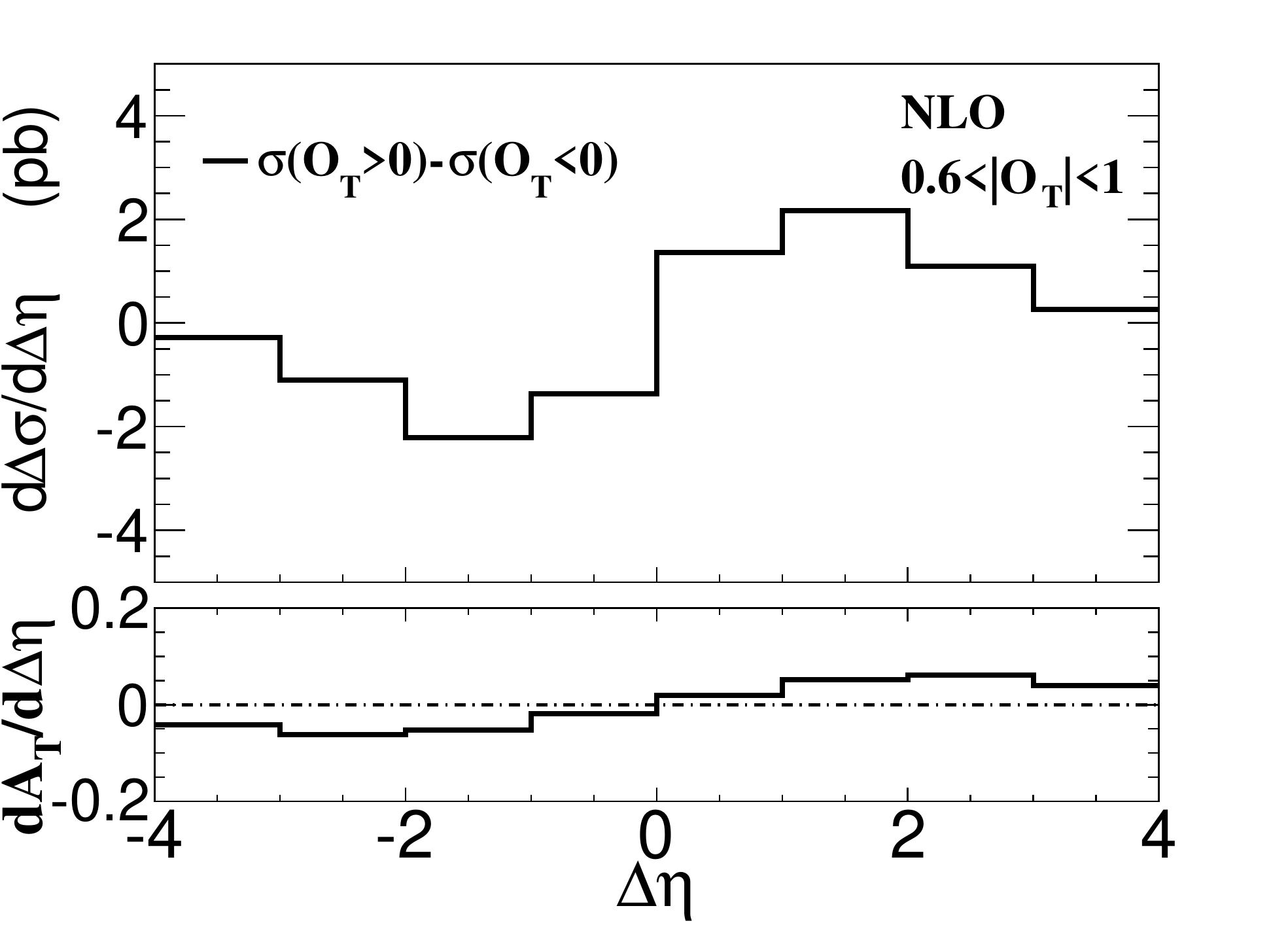}
\includegraphics[width=.45\textwidth]{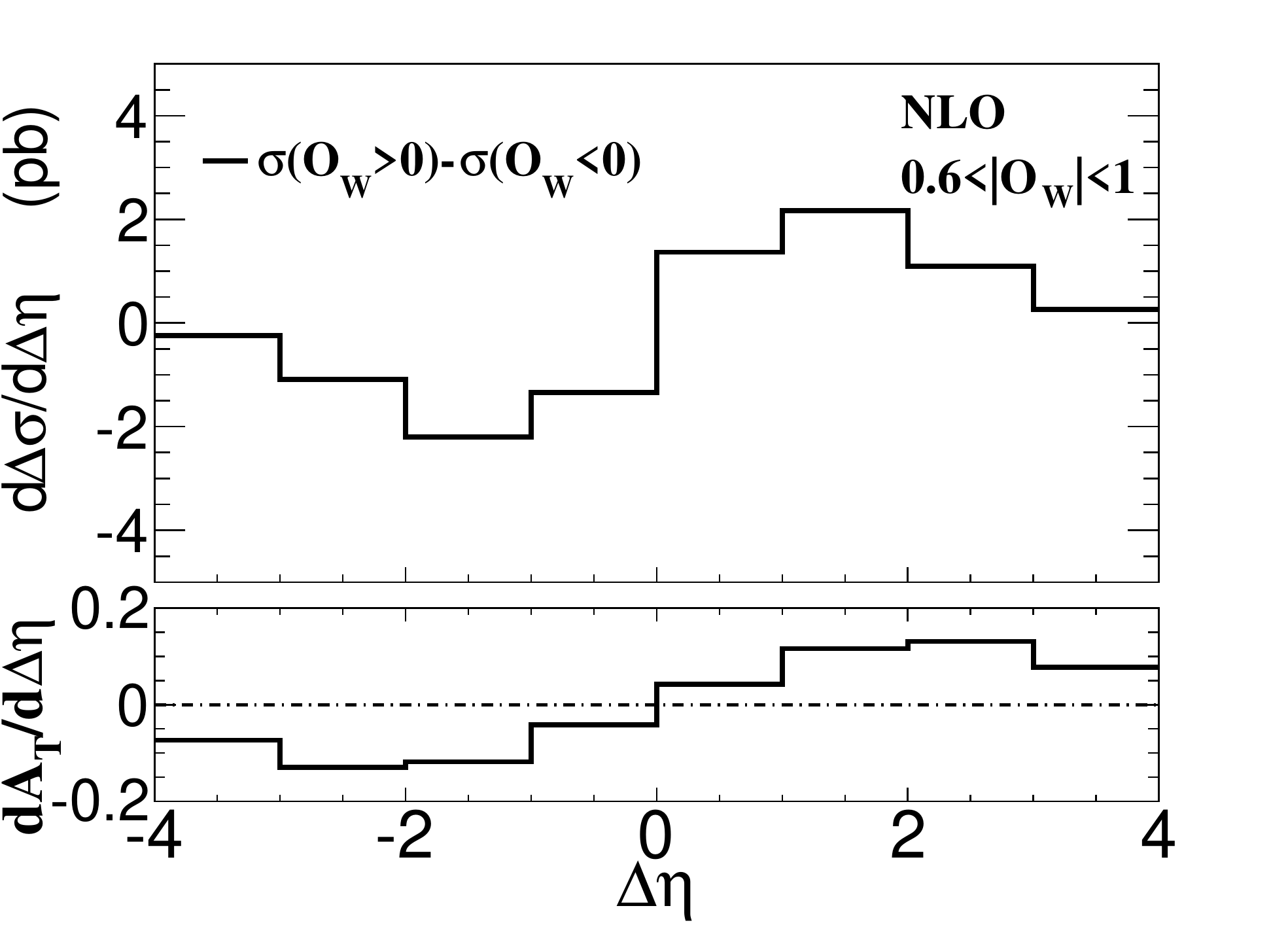}
\caption{ The SM-NLO asymmetries  ${\cal O}_T$ (left) and ${\cal O}_W$ (right)  with the additional cuts $0.6<|{\cal O}_{T} | < 1$ and $0.6<|{\cal O}_{W} | < 1 $ as discussed in the text for 13 ~TeV.
\label{fig:NLO_detatw}}
\end{figure}

The corresponding distributions for $\Delta \sigma$ with respect to $\Delta \eta$ are shown in Fig.~\ref{fig:NLO_detatw}. The two operators,  ${\cal O}_{T}$ and ${\cal O}_{W}$, lead to very similar differential distributions $d\Delta\sigma/d\Delta\eta$ but $\mathcal{O}_W$ produces a larger asymmetry, as explained above. We also find that the NLO-SM asymmetries at 13 TeV have very similar size to those at 8 TeV when we apply the cut $p_T(W^+)>30$ GeV. 

\begin{figure}[thb]
\includegraphics[width=.45\textwidth]{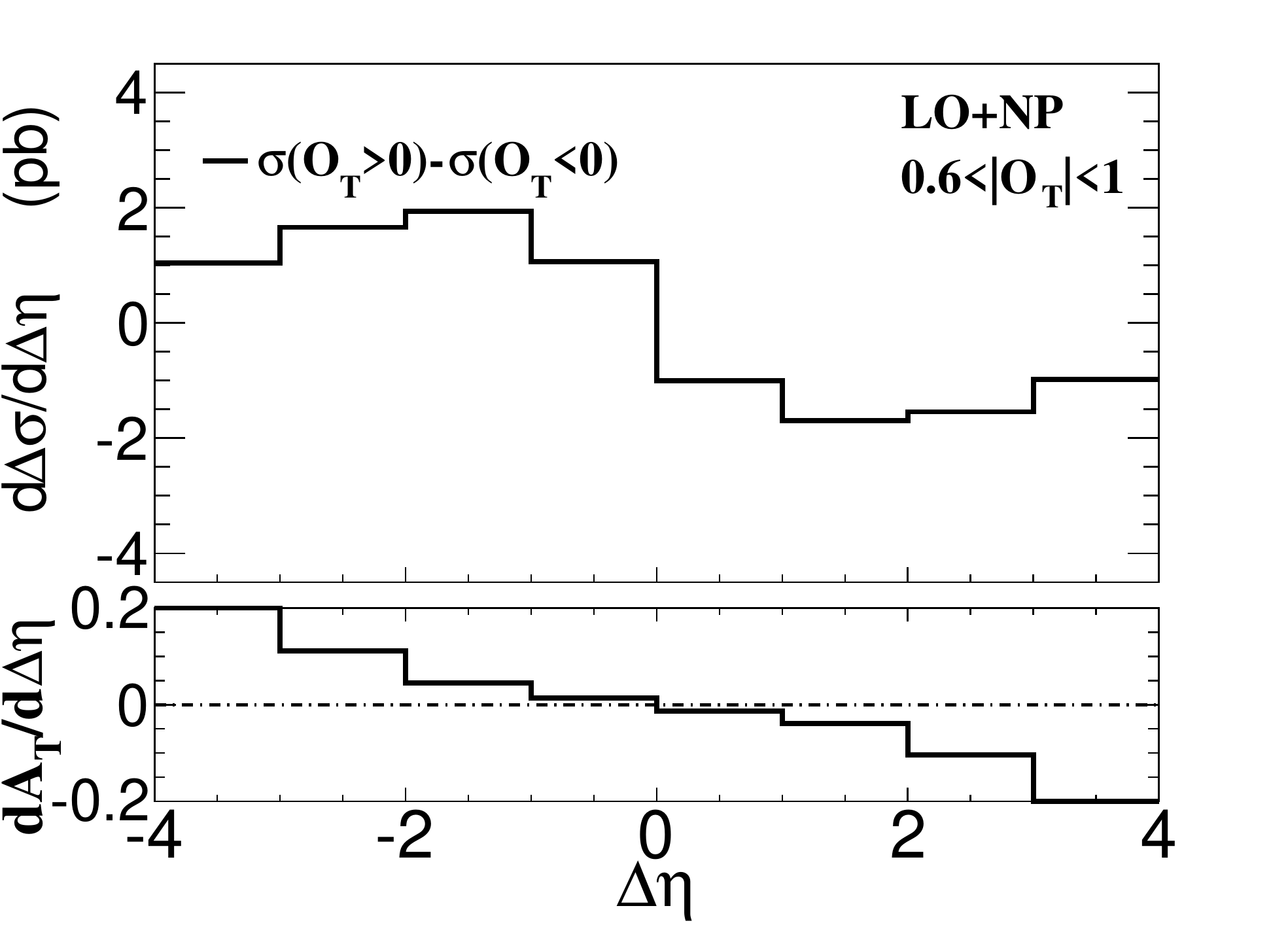}
\caption{ The SM-LO+NP asymmetry  $O_{T}$  with an extra cut  $0.6<|{\cal O}_{T} | < 1$ at 13~TeV. For this plot we have used ${\rm ~Im}(g_{12})/\Lambda^4= 10^{-8}$~GeV$^{-4}$.
\label{fig:NP_detat}}
\end{figure}

\begin{figure}[thb]
\includegraphics[width=.65\textwidth]{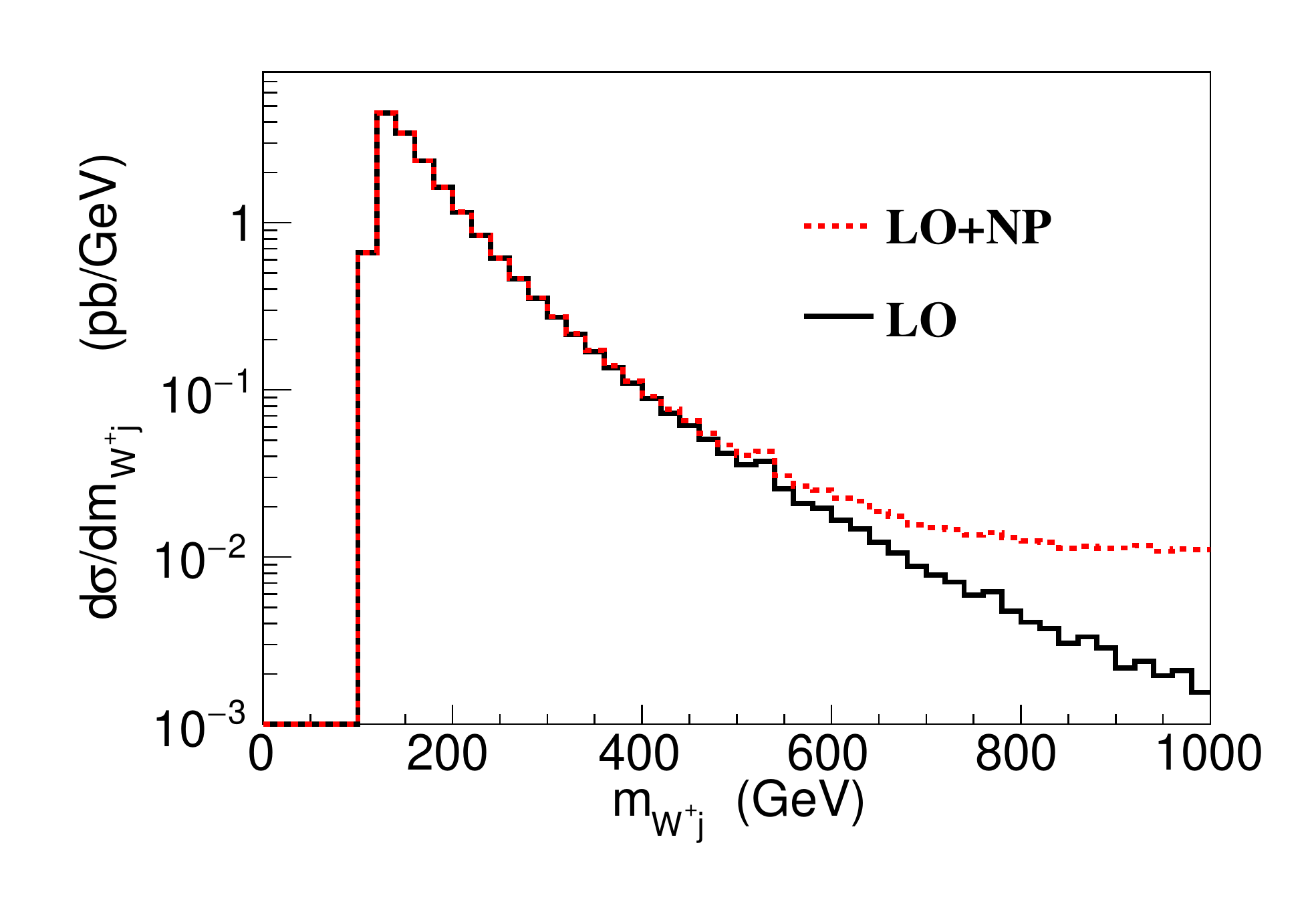}
\caption{The SM and SM+NP  $M_{Wj}$ distributions for illustrative parameters  ${\rm Im}(g_{12})/\Lambda^4=1.23\times 10^{-10}$~GeV$^{-4}$. 
\label{fig:mwj_NP_LO}}
\end{figure}

Having established the size of the asymmetries produced by NLO QCD effects we turn our attention to those arising from the CP violating inference between the LO-SM and the NP amplitudes. These asymmetries will scale with ${\rm Im}(g_{12})/\Lambda^4$ as is obvious from Eq.~\ref{opweuse}. A first crude estimate for an observable asymmetry is one of similar size to that produced by NLO QCD. Choosing  $\Lambda=1$~TeV, as is common for LHC studies, requires ${\rm Im}(g_{12}) \sim 10^4$ indicating right away that CP violating new physics is not likely to be a significant complication in trying to measure the NLO QCD phase. We illustrate the resulting asymmetry in Fig.~\ref{fig:NP_detat} for the case of $\mathcal{O}_T$ (which is the same as $\mathcal{O}_W$  at LO). To keep most of the events and enhance the asymmetry we apply the additional cuts $0.6<|\mathcal{O}_{T(W)}|<1$ and $\Delta \eta >1$. Comparing Figures~\ref{fig:NLO_detatw}~and~\ref{fig:NP_detat} we see very similar shapes for NLO QCD and CP violation, with the only discernible difference being that the latter is relatively larger at large $\Delta\eta$. The inverted shape about $\Delta\eta=0$ is simply due to the sign choice for ${\rm Im}(g_{12})$.

The integrated asymmetries are summarized in Table~\ref{tab:Aotow} confirming  that the LHC is only sensitive to a new physics scale near a few hundred GeV in this case, much lower than what is excluded by low energy FCNC constraints. 

To understand why this sensitivity is so low, we recall that the SM $W$~plus~jet cross section peaks strongly near $m_{Wj}=100$~GeV and has dropped by a factor of 5 by 200~GeV. Therefore, the interference with the NP mostly samples the low $m_{Wj}$ region where it is not enhanced by the growth with energy that characterizes non-renormalizable dimension eight amplitudes. This is illustrated in Figure~\ref{fig:mwj_NP_LO}. This suggests, in turn, that the LHC can place stronger constraints on this operator from its quadratic enhancement of the cross-section that appears for large values of $m_{Wj}$ as seen in the figure. To quantify this statement, we compute the total cross-section (LO-SM + NP)  keeping only the imaginary part of the new coupling $g_{12}$ in the region $500\leq m_{Wj} \leq 800$~GeV for a new physics scale $\Lambda=1$~TeV. The contribution from the SM-NP interference to the total cross section vanishes for an imaginary $g_{12}$, and we find 
\begin{eqnarray}
\sigma\approx \left[\sigma_{SM}+ 1.29\times 10^{-4}\  {\rm Im}(g_{12})^2 \left(\frac{1~{\rm TeV}}{\Lambda}\right)^8\right] {\rm ~pb},
\label{dev}
\end{eqnarray}
where $\sigma_{SM_{\rm LO}}= 4.68$~pb. Assuming that this cross-section can eventually be measured and that statistical error will dominate the measurement, 100~fb$^{-1}$ will provide the 95\% c.l. constraint (since the deviation from the SM in Eq.~(\ref{dev}) can only be upwards),
\begin{eqnarray}
{\rm Im}(g_{12}) \left(\frac{1~{\rm TeV}}{\Lambda}\right)^4 \lsim 14.3.
\end{eqnarray}
Alternatively, with ${\rm Im}(g_{12})\sim 1$, this process will be sensitive to a new physics scale $\Lambda \lsim 514$~GeV, not quite reaching the level implied by rare kaon decay constraints.

\begin{table}
\begin{tabular}{c|c|c}
\hline\hline
 $\mathcal{A}_{T,NLO} $ & $ \mathcal{A}_{W,NLO}$ &   $\mathcal{A}_{T,LO+NP}$ \\
 \hline
 5.2\% & 11.2\% & -8.4\% $ \times \mathrm{Im}(g_{12}) \left(\frac{100 \mathrm{GeV}} {\Lambda}\right)^4 $
 \\
 \hline\hline
\end{tabular}
\caption{\label{tab:Aotow} The total asymmetries for $\mathcal{O}_{T}$ and $\mathcal{O}_{W}$ with the cut $0.6<|\mathcal{O}_{T(W)}|<1$ and $\Delta \eta >1$~.}
\end{table}

\begin{figure}[thb]
\includegraphics[width=.45\textwidth]{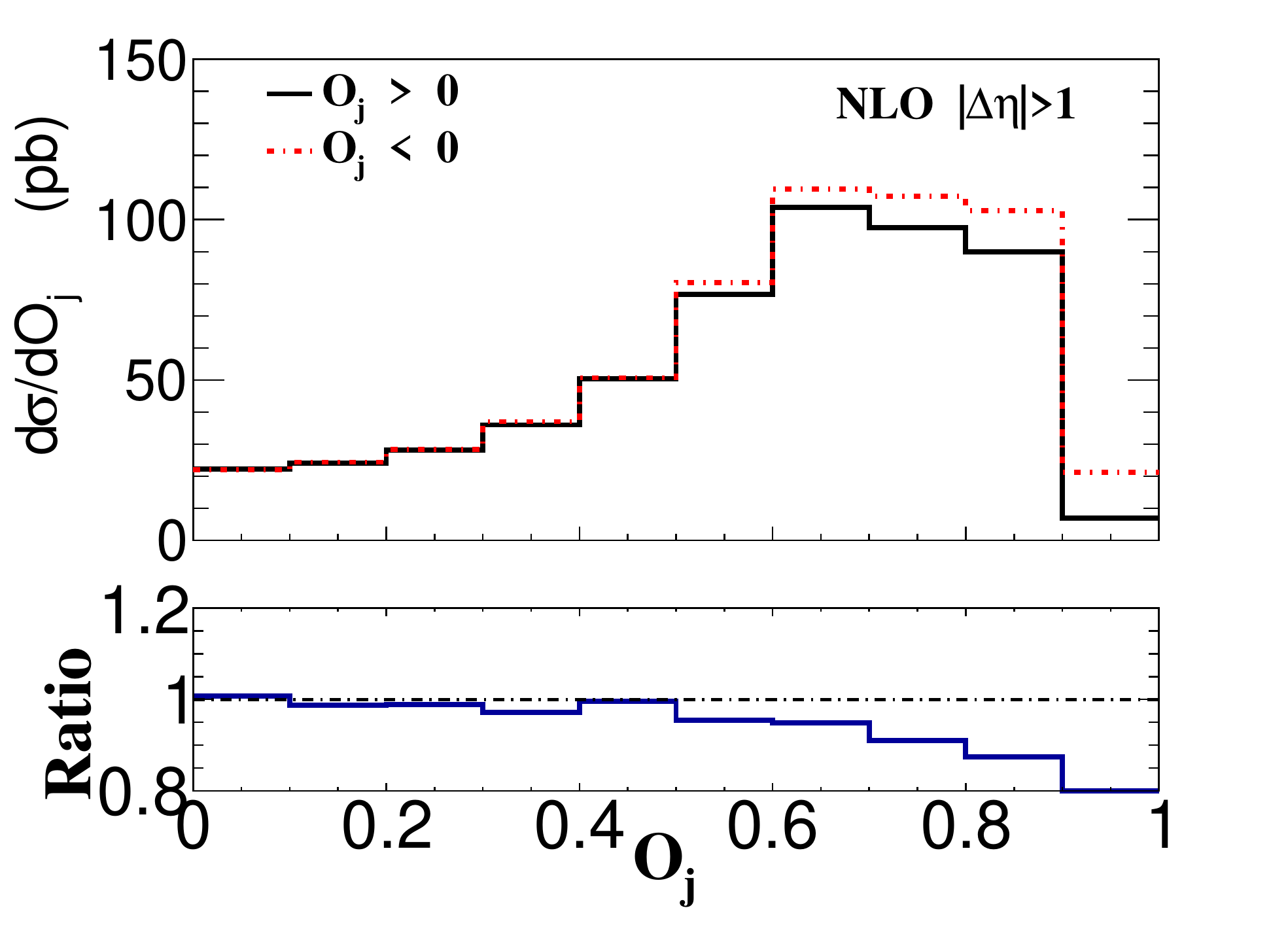}
\includegraphics[width=.45\textwidth]{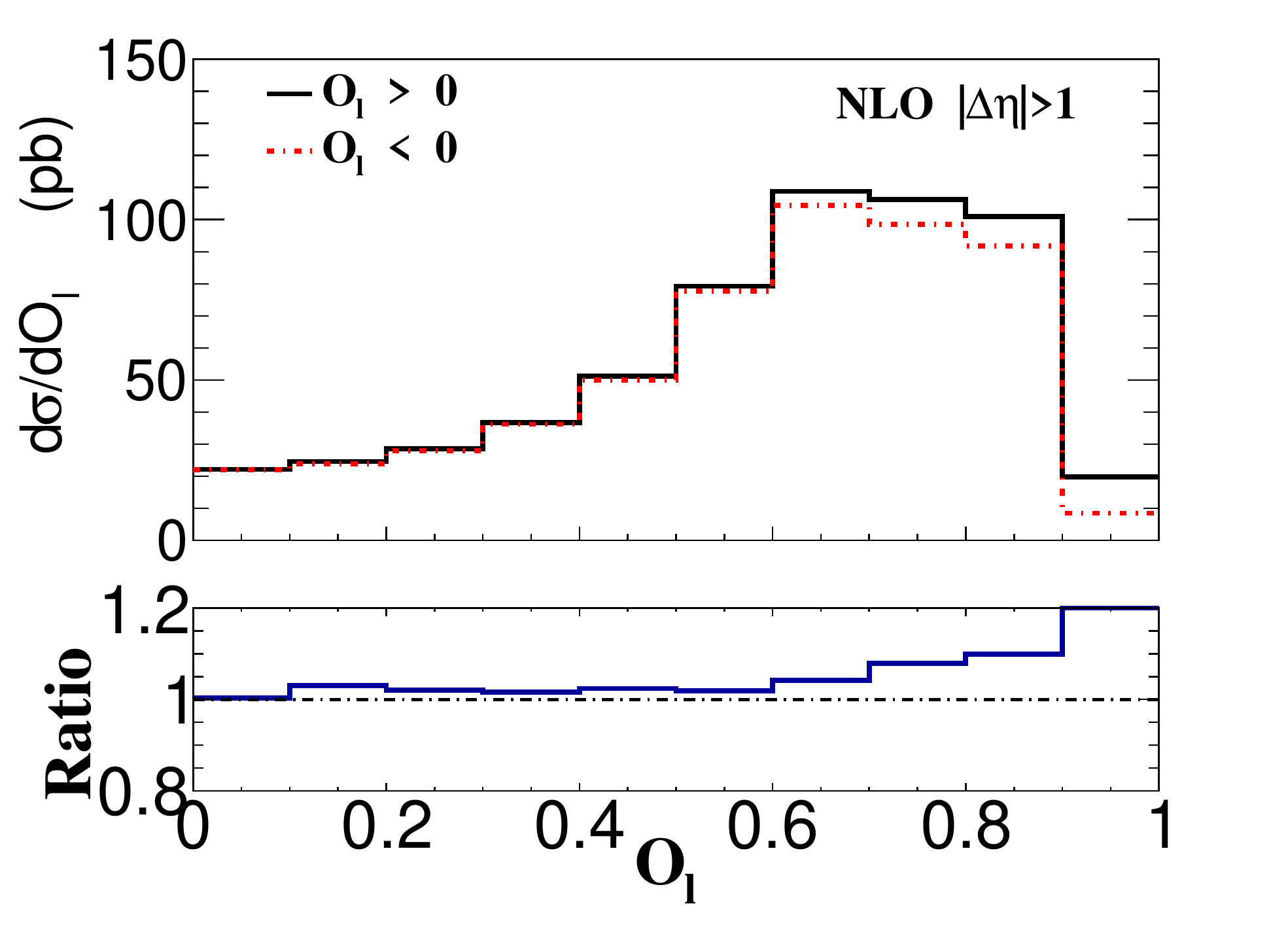}
\includegraphics[width=.45\textwidth]{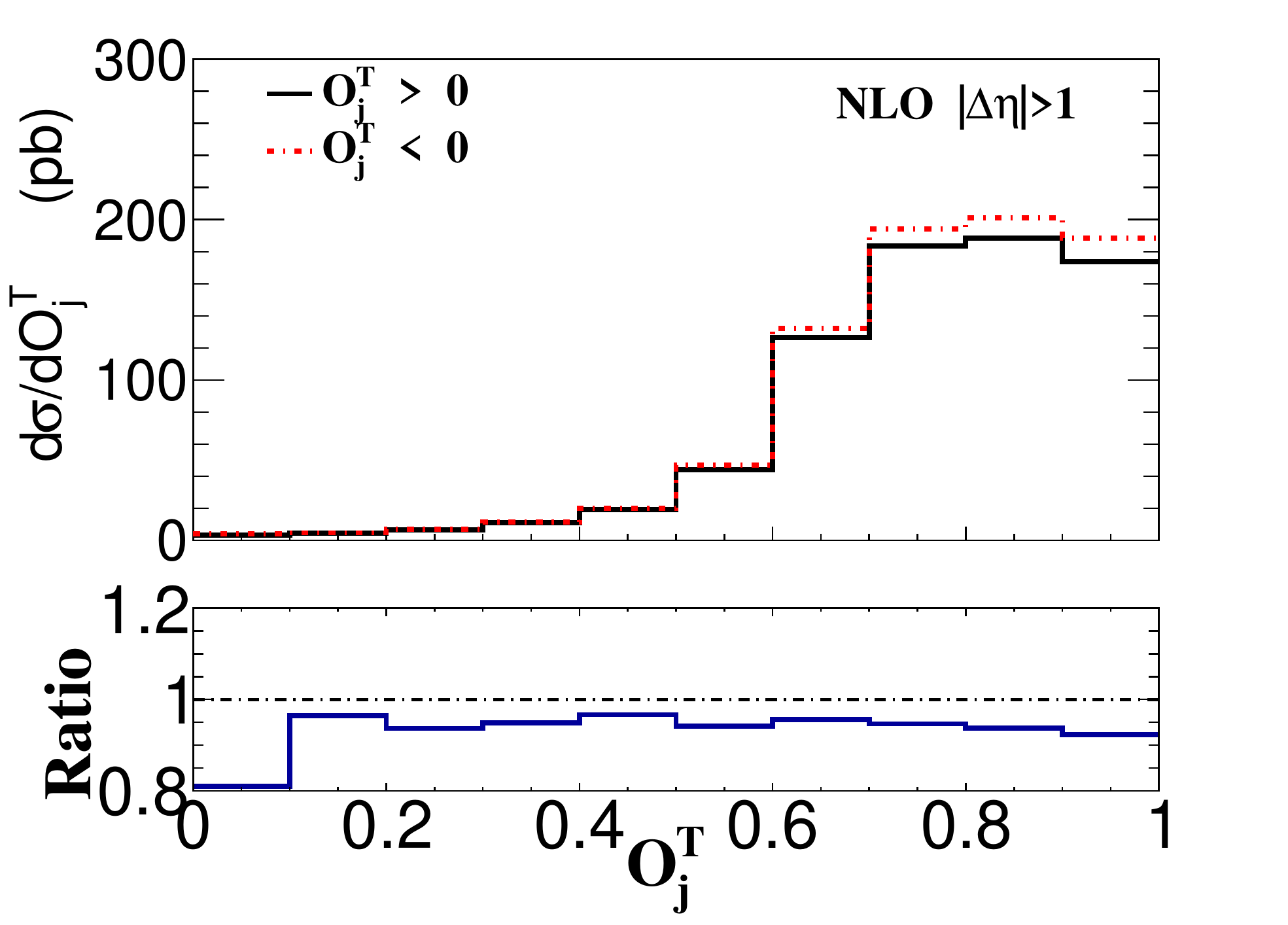}
\includegraphics[width=.45\textwidth]{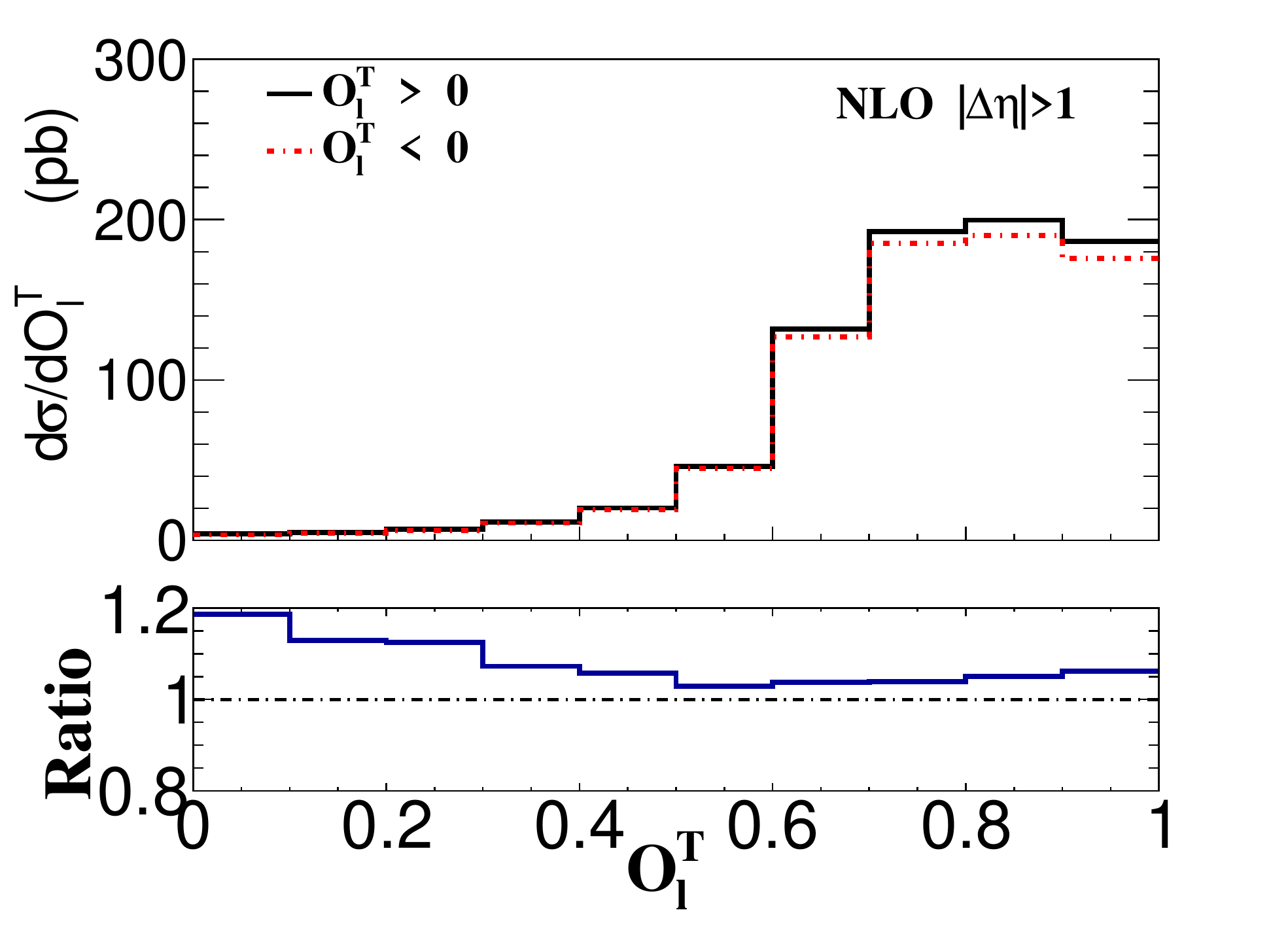}
\caption{ Comparison between the distributions with positive and negative values of  $\mathcal{O}_{j}$, $\mathcal{O}_{\ell}$, $\mathcal{O}_{j}^{(T)}$ and $\mathcal{O}_{\ell}^{(T)}$ for the SM-NLO. 
\label{fig:NLOxtjl}}
\end{figure}

\begin{figure}[thb]
\includegraphics[width=.45\textwidth]{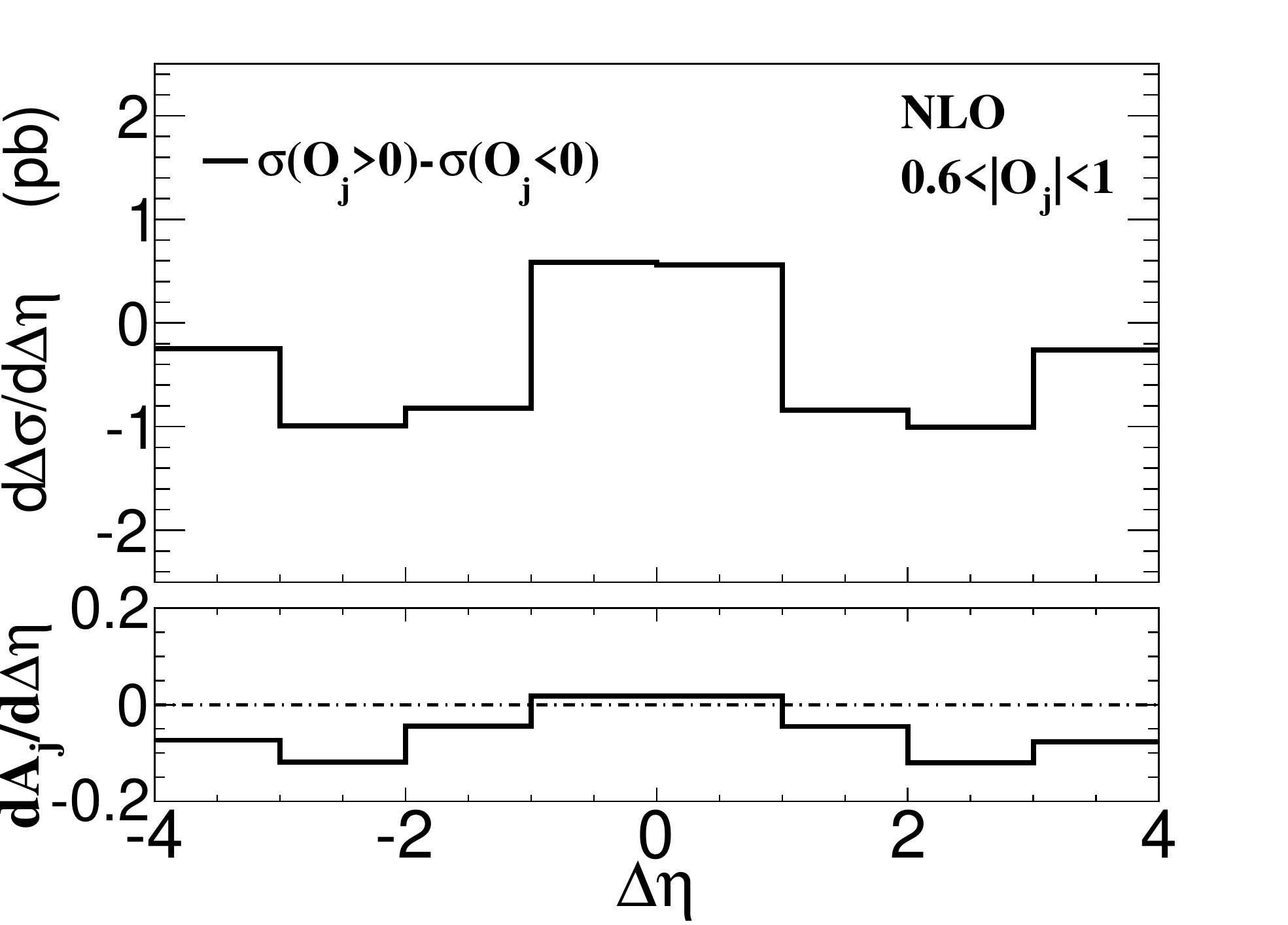}
\includegraphics[width=.45\textwidth]{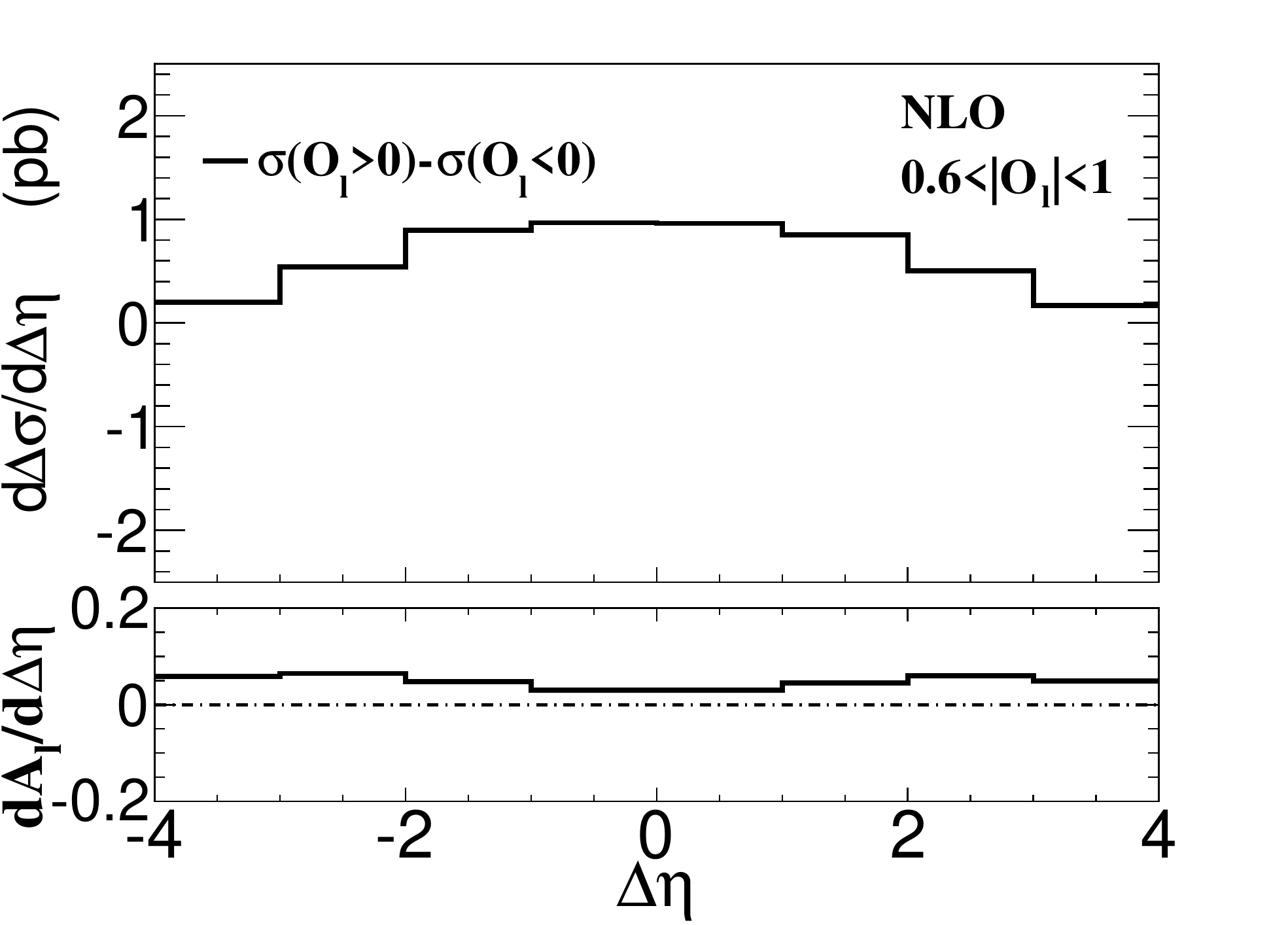}
\includegraphics[width=.45\textwidth]{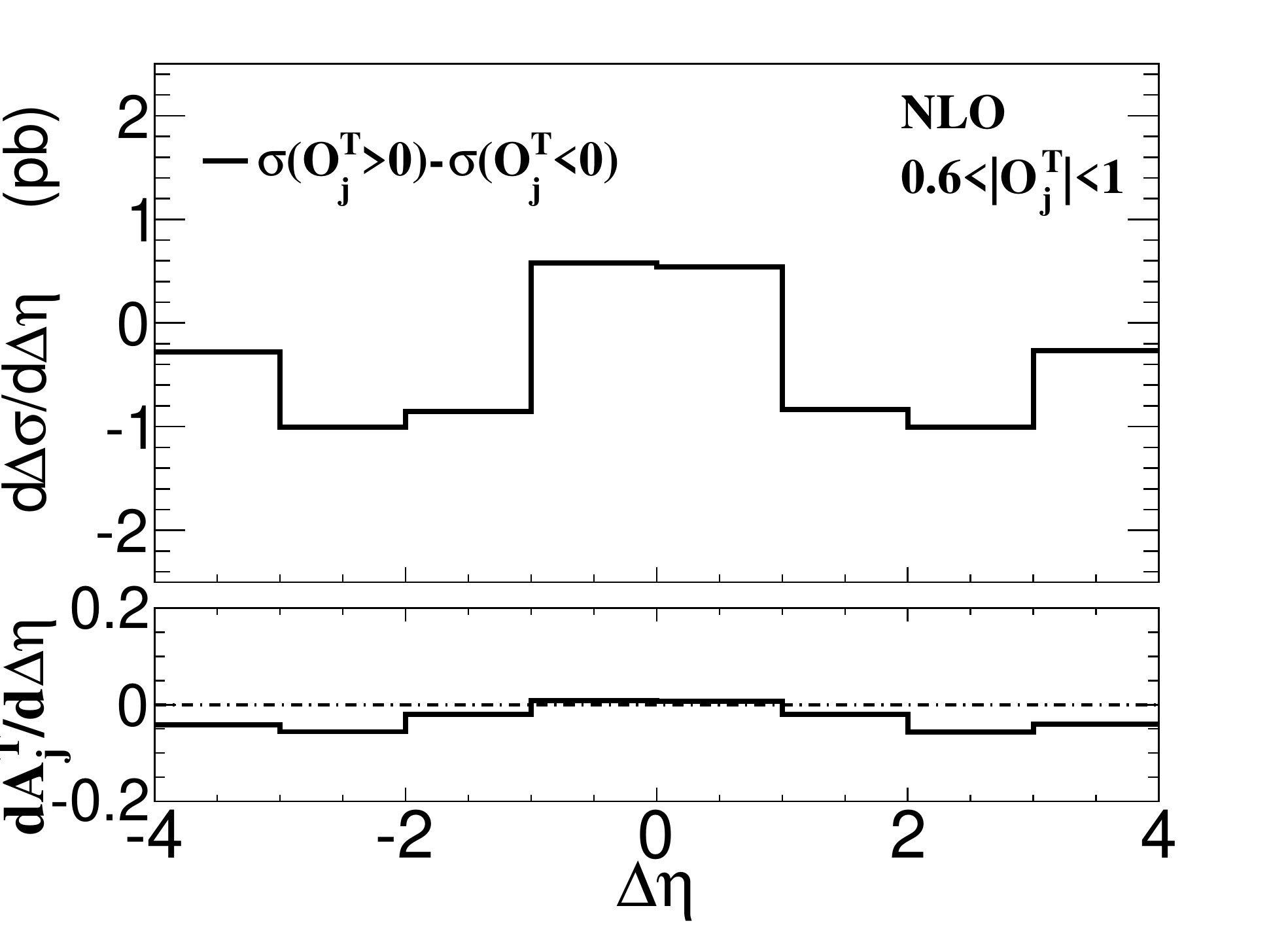}
\includegraphics[width=.45\textwidth]{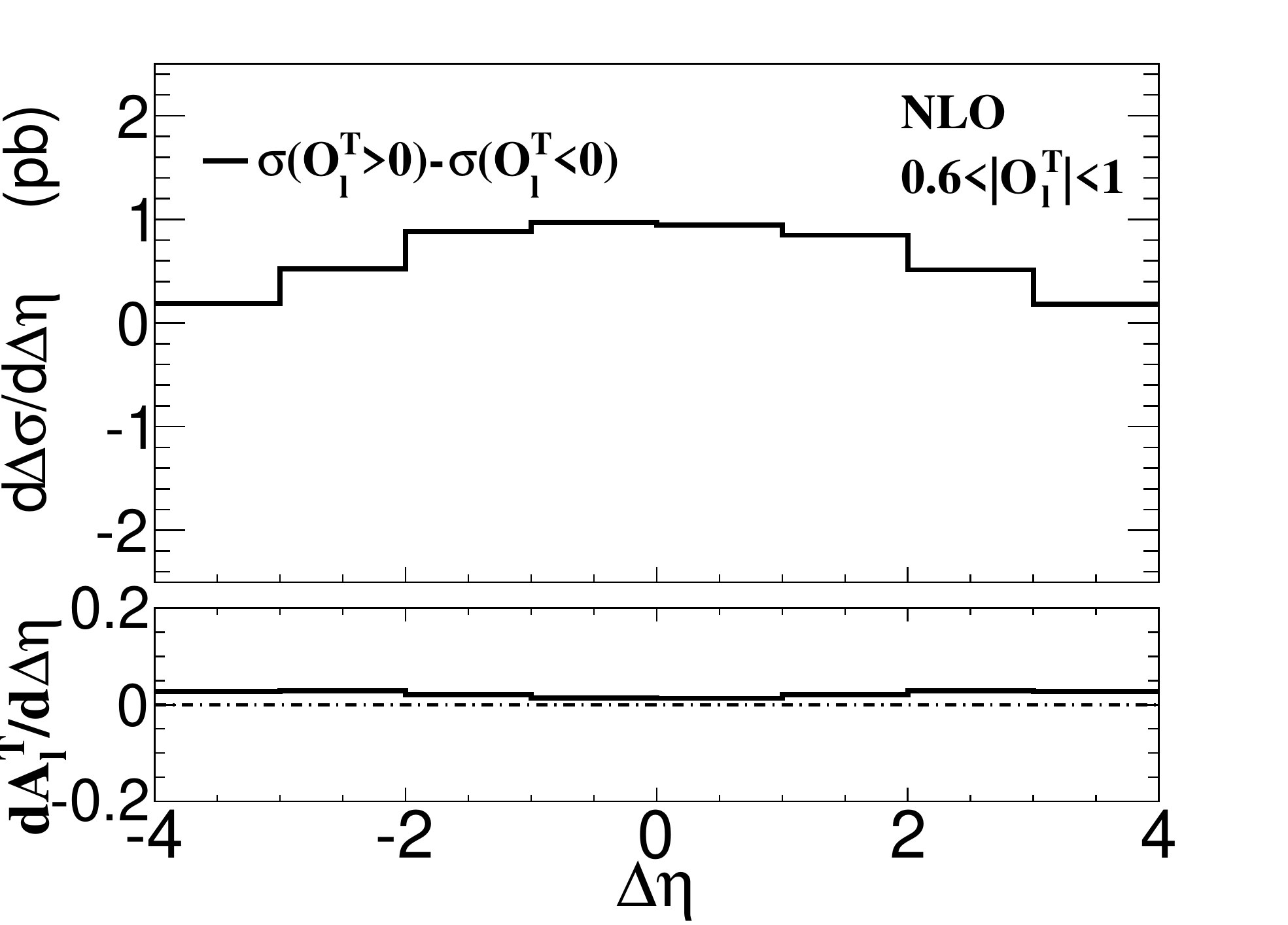}
\caption{ $\Delta \eta $ distributions for $\Delta \sigma$ and the asymmetries $\mathcal{A}$  for the SM-NLO results~.
\label{fig:NLOdeta_jl}}
\end{figure}

\begin{figure}[thb]
\includegraphics[width=.45\textwidth]{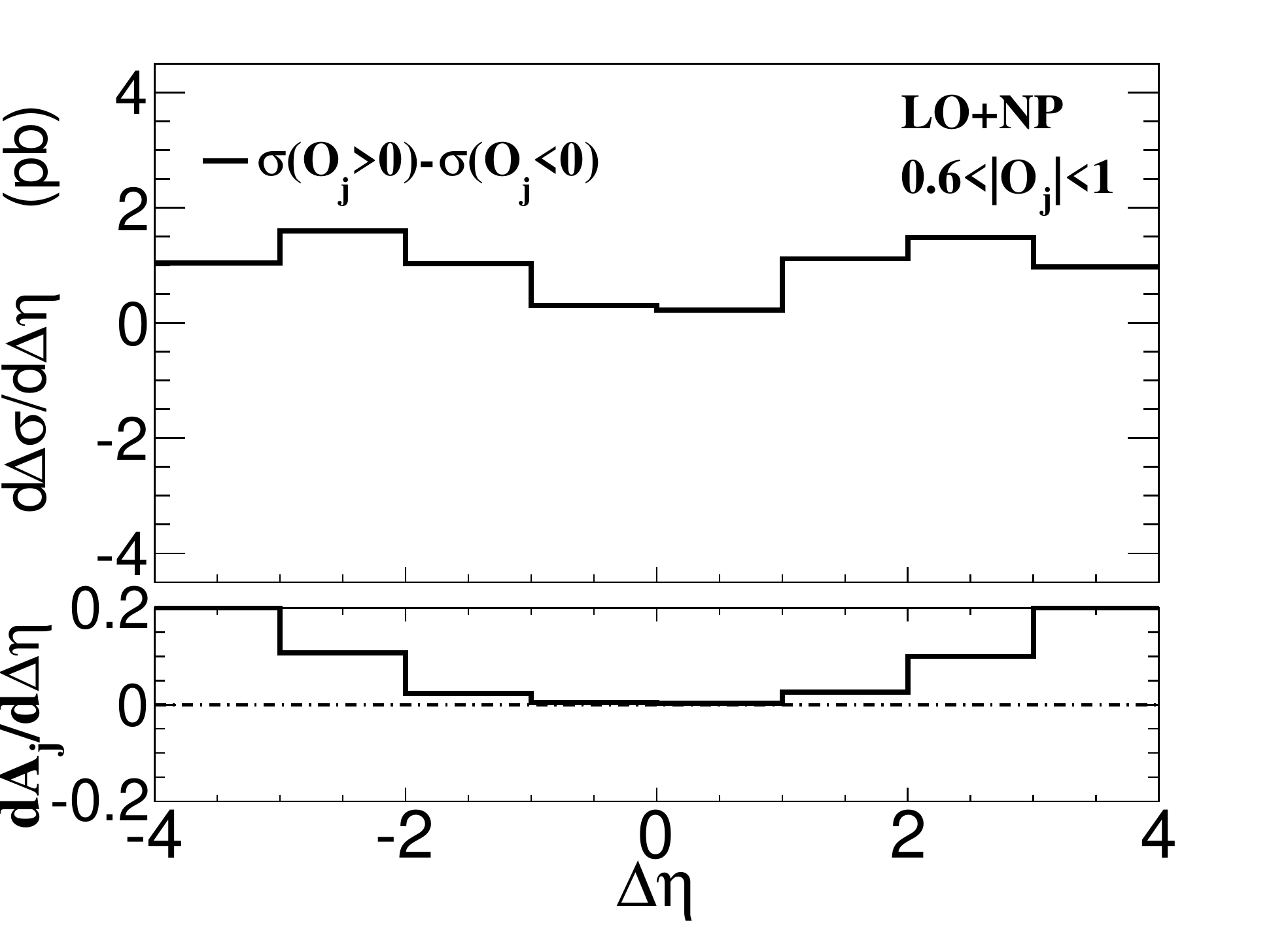}
\includegraphics[width=.45\textwidth]{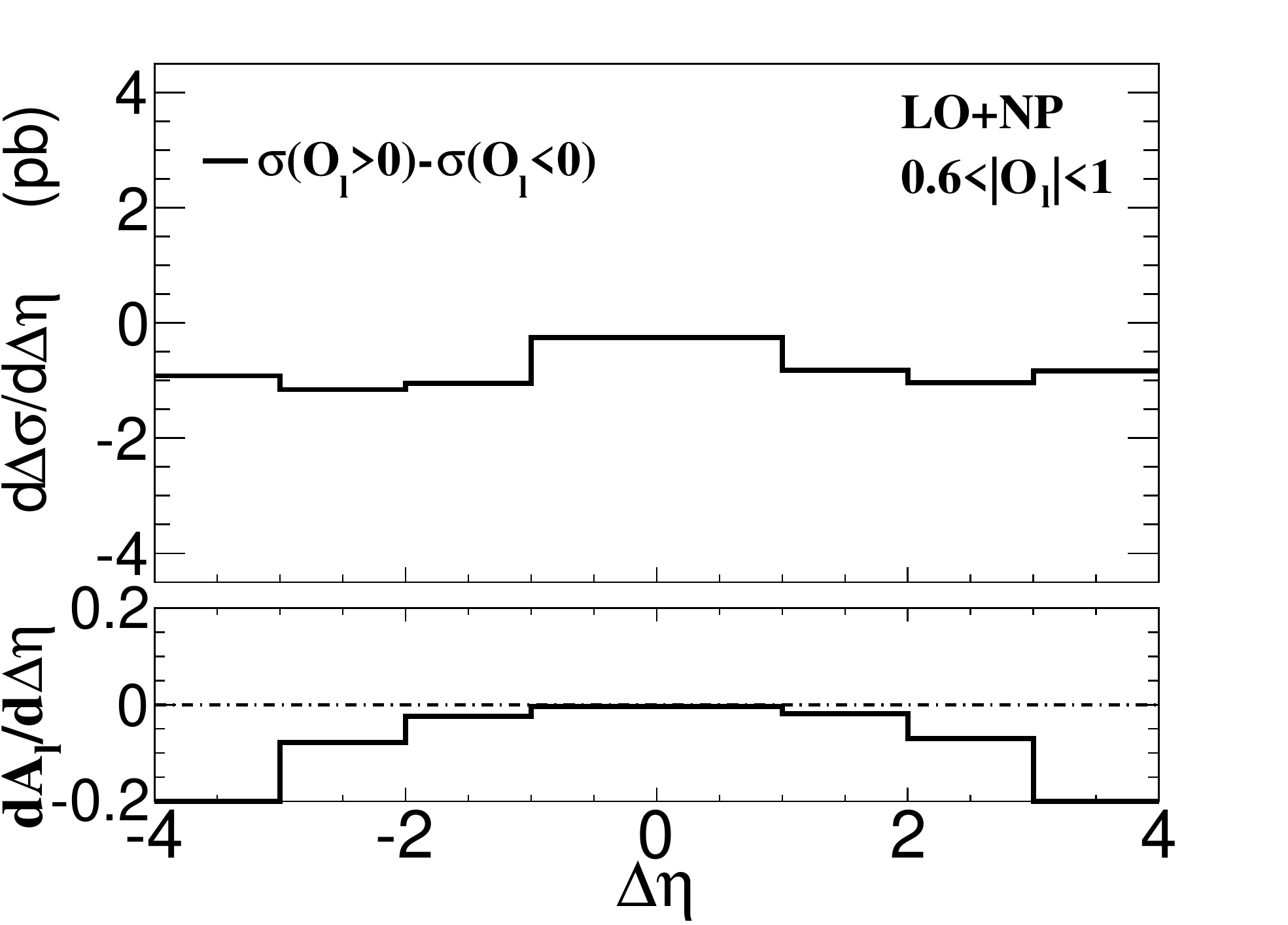}
\caption{ $\Delta \eta $ distributions for  the asymmetries for the SM-LO+NP results. 
\label{fig:NPdeta_jl}}
\end{figure}

For completeness we illustrate the T-odd correlations defined in Eq.~(\ref{eq:tripprods}), which explicitly exhibit the symmetry of the initial $pp$ state at LHC. Fig.~\ref{fig:NLOxtjl} shows the distributions with respect to $\mathcal{O}_{j\ell}$ and $\mathcal{O}^T_{j\ell}$ which can be compared with those of Figure~\ref{fig:NP_detat}. There appears to be no clear advantage for either method of adapting $\mathcal{O}^W$ for an LHC measurement: a cut on $\Delta \eta$ vs a cut on $|\Delta\eta |$ supplemented with the additional factor that defines $\mathcal{O}_{j\ell}$. Comparing Figures~\ref{fig:NLOdeta_jl}~and~\ref{fig:NLO_detatw} we see that the additional factor that defines $\mathcal{O}_{j\ell}$ symmetrizes the distributions as it was designed to do, but again does not appear to provide any advantage for the extraction of the NLO-QCD phase.
The integrated asymmetries are presented in table~\ref{tab:Aojl}~, once again producing results that are comparable to those of Table~\ref{tab:Aotow}~.
\begin{table}
\begin{tabular}{c|c|c|c|c|c}
\hline\hline
\multirow{2}{*}{ $\mathcal{A}_{j,NLO} $ }   & 
\multirow{2}{*}{ $ \mathcal{A}_{\ell,NLO}$}  & 
\multirow{2}{*}{ $\mathcal{A}^{T}_{j,NLO} $} &
\multirow{2}{*}{ $\mathcal{A}^{T}_{\ell,NLO}$ }  & 
 $\mathcal{A}_{j,LO+NP}$ & $\mathcal{A}_{\ell,LO+NP}$ \\
 \cline{5-6}
   & & & & \multicolumn{2}{c}{$\mathrm{Im}(g_{12}) \left(\frac{100 \mathrm{GeV}} {\Lambda}\right)^4$}
 \\
 
 \hline
 -6.7\% & 5.1\% & -3.0\% & 2.3\%  & 6.4\%  & -6.0\%
 \\
 \hline\hline
\end{tabular}
\caption{\label{tab:Aojl} The total asymmetries for $\mathcal{O}_{j(\ell)}^{(T)}$ with the cut $0.6<|\mathcal{O}_{j(\ell)}^{(T)}|<1$ and $|\Delta \eta | >1 $~.}
\end{table}

\section{Summary and Conclusions}

We have studied T-odd correlations in the process $pp \to Wj$ at the LHC for 13~TeV. These correlations can probe the phase of the NLO QCD corrections as discussed in Reference~\cite{Frederix:2014cba} for 8~TeV. The asymmetries we show in Figure~\ref{fig:NLO_otow} are very similar to the ones found at 8~TeV in Figure~\ref{fig:NLO_comp} and the cross-section for $Wj$ production at 8~TeV is only about a factor of 2 smaller than at 13~TeV. Our conclusion is therefore that this asymmetry can also be measured at the LHC at 13 TeV.

We have then investigated the possibility of generating the T-odd correlations through interference between the LO-SM and CP violating new physics. We have parametrized the new physics in terms of an effective Lagrangian that respects the symmetries of the SM, and have studied the conditions that the operators in this Lagrangian need to satisfy to generate a sizeable CP-odd interference with the SM. We found that operators  which can generate such interference do not occur at dimension six and first appear at dimension eight. Moreover, we found that the operator necessarily induces FCNC and its coefficient is constrained by rare meson decay.

We computed the T-odd asymmetries from this NP operator and required them to have similar size to those from NLO QCD to find that a very low NP scale (a few 100 GeV) would be needed. Since this scale is much lower than the constraints from rare meson decay, we conclude that CP violating NP is not a relevant background when measuring the NLO QCD phase.

Finally we estimate the constraints that can be placed on the NP at the LHC in two different ways. From deviations in the total cross-section for $500 \leq m_{Wj} \leq 800$~GeV we found 
${\rm Im}(g_{12}) \lsim 14.3$ for a NP scale of 1~TeV.

Assuming that the QCD asymmetries are precisely known and measured, the statistical error in that measurement would also constrain the NP. A simple estimate can be obtained as follows: 
a QCD integrated asymmetry of typical size, $\sim 5\%$ in a cross-section $\sigma\sim 70$~pb as in Table~\ref{tab:Aojl}, can be measured with 100 fb$^{-1}$ with a $1\sigma$ statistical error 
\begin{eqnarray}
\delta \mathcal{A} = \sqrt{(1-\mathcal{A}^2)/N} \sim  4\times 10^{-4}
\end{eqnarray}
Requiring the NP asymmetry to be below this level results in the constraint
\begin{eqnarray}
{\rm Im}(g_{12}) \left(\frac{1~{\rm TeV}}{\Lambda}\right)^4 \lsim  67,
\end{eqnarray}
roughly four times worse than can be obtained from considering the total cross-section.

Our findings suggest that any NP CP violating effects will not pollute a measurement of the NLO QCD phases at the LHC, and in fact, that they can only be probed with high statistics and only as corrections to precisely known QCD phases.

\begin{acknowledgments}

We are grateful to Celine Degrande and Olivier Mattelaer for useful discussions and Hiroshi Yokoya for clarifications on Ref.~\cite{Frederix:2014cba}. G.V. thanks the theory group at CERN for their hospitality and partial support while  this work was completed. The work of H.L.  was supported in part by the ARC Centre of Excellence for Particle Physics at the Terascale.

\end{acknowledgments}


\begin{thebibliography}{999}
   
\bibitem{Frederix:2014cba} 
  R.~Frederix, K.~Hagiwara, T.~Yamada and H.~Yokoya,
  Phys.\ Rev.\ Lett.\  {\bf 113}, 152001 (2014)
  doi:10.1103/PhysRevLett.113.152001
  [arXiv:1407.1016 [hep-ph]].
  
  
\bibitem{Aad:2014qxa} 
  G.~Aad {\it et al.} [ATLAS Collaboration],
  Eur.\ Phys.\ J.\ C {\bf 75}, no. 2, 82 (2015)
  doi:10.1140/epjc/s10052-015-3262-7
  [arXiv:1409.8639 [hep-ex]].
  
\bibitem{Khachatryan:2016fue} 
  V.~Khachatryan {\it et al.} [CMS Collaboration],
  [arXiv:1610.04222 [hep-ex]].
  
\bibitem{Sirunyan:2017wgx} 
  A.~M.~Sirunyan {\it et al.} [CMS Collaboration],
  arXiv:1707.05979 [hep-ex].
  
\bibitem{Dawson:1995wg} 
  S.~Dawson and G.~Valencia,
  Phys.\ Rev.\ D {\bf 52}, 2717 (1995)
  doi:10.1103/PhysRevD.52.2717
  [hep-ph/9504209].
  
\bibitem{Hagiwara:2006qe} 
  K.~Hagiwara, K.~i.~Hikasa and H.~Yokoya,
  Phys.\ Rev.\ Lett.\  {\bf 97}, 221802 (2006)
  doi:10.1103/PhysRevLett.97.221802
  [hep-ph/0604208].

\bibitem{Buchmuller:1985jz} 
  W.~Buchmuller and D.~Wyler,
  Nucl.\ Phys.\ B {\bf 268}, 621 (1986).

\bibitem{Grzadkowski:2010es} 
  B.~Grzadkowski, M.~Iskrzynski, M.~Misiak and J.~Rosiek,
  JHEP {\bf 1010}, 085 (2010)
  [arXiv:1008.4884 [hep-ph]].
  

\bibitem{Alonso:2012px} 
  R.~Alonso, M.~B.~Gavela, L.~Merlo, S.~Rigolin and J.~Yepes,
  Phys.\ Lett.\ B {\bf 722}, 330 (2013)
  Erratum: [Phys.\ Lett.\ B {\bf 726}, 926 (2013)]
  doi:10.1016/j.physletb.2013.04.037, 10.1016/j.physletb.2013.09.028
  [arXiv:1212.3305 [hep-ph]].
  
\bibitem{Buchalla:2013rka} 
  G.~Buchalla, O.~Cata and C.~Krause,
  Nucl.\ Phys.\ B {\bf 880}, 552 (2014)
  Erratum: [Nucl.\ Phys.\ B {\bf 913}, 475 (2016)]
  doi:10.1016/j.nuclphysb.2016.09.010, 10.1016/j.nuclphysb.2014.01.018
  [arXiv:1307.5017 [hep-ph]].

\bibitem{Gavela:2014vra} 
  M.~B.~Gavela, J.~Gonzalez-Fraile, M.~C.~Gonzalez-Garcia, L.~Merlo, S.~Rigolin and J.~Yepes,
  JHEP {\bf 1410}, 044 (2014)
  doi:10.1007/JHEP10(2014)044
  [arXiv:1406.6367 [hep-ph]].


\bibitem{Buras:1992uf} 
A.~J.~Buras and M.~K.~Harlander,
  Adv.\ Ser.\ Direct.\ High Energy Phys.\  {\bf 10}, 58 (1992).

  
\bibitem{Inami:1980fz} 
  T.~Inami and C.~S.~Lim,
  Prog.\ Theor.\ Phys.\  {\bf 65}, 297 (1981)
  Erratum: [Prog.\ Theor.\ Phys.\  {\bf 65}, 1772 (1981)].
  
  \bibitem{expres}
  S.~Adler {\it et al.} [E787 Collaboration],
  Phys.\ Rev.\ Lett.\  {\bf 88}, 041803 (2002)
  doi:10.1103/PhysRevLett.88.041803
  [hep-ex/0111091];
  V.~V.~Anisimovsky {\it et al.} [E949 Collaboration],
  Phys.\ Rev.\ Lett.\  {\bf 93}, 031801 (2004)
  doi:10.1103/PhysRevLett.93.031801
  [hep-ex/0403036].

  
\bibitem{Olive:2016xmw} 
  C.~Patrignani {\it et al.} [Particle Data Group],
  Chin.\ Phys.\ C {\bf 40}, no. 10, 100001 (2016).
  doi:10.1088/1674-1137/40/10/100001


\bibitem{Descotes-Genon:2015uva} 
  S.~Descotes-Genon, L.~Hofer, J.~Matias and J.~Virto,
  JHEP {\bf 1606}, 092 (2016)
  doi:10.1007/JHEP06(2016)092
  [arXiv:1510.04239 [hep-ph]].
  
  
%
  
\bibitem{Dawson:2013owa} 
  S.~Dawson, S.~K.~Gupta and G.~Valencia,
  Phys.\ Rev.\ D {\bf 88}, no. 3, 035008 (2013)
  doi:10.1103/PhysRevD.88.035008
  [arXiv:1304.3514 [hep-ph]].
  
  
\bibitem{Alwall:2014hca} 
 J.~Alwall {\it et al.},
 JHEP {\bf 1407}, 079 (2014)
 [arXiv:1405.0301 [hep-ph]].
  
\bibitem{Christensen:2008py} 
 N.~D.~Christensen and C.~Duhr,
 Comput.\ Phys.\ Commun.\  {\bf 180}, 1614 (2009)
 [arXiv:0806.4194 [hep-ph]].

\bibitem{Degrande:2011ua} 
 C.~Degrande, C.~Duhr, B.~Fuks, D.~Grellscheid, O.~Mattelaer and T.~Reiter,
 Comput.\ Phys.\ Commun.\  {\bf 183}, 1201 (2012)
 [arXiv:1108.2040 [hep-ph]].

 
\bibitem{Dulat:2015mca} 
  S.~Dulat {\it et al.},
  Phys.\ Rev.\ D {\bf 93}, no. 3, 033006 (2016)
  doi:10.1103/PhysRevD.93.033006
  [arXiv:1506.07443 [hep-ph]].
  
   
\end{thebibliography}
\end{document}